\documentclass[preprint,12pt,3p]{elsarticle}

\usepackage{amssymb}
\usepackage{lineno,hyperref}
\usepackage{amsmath}
\usepackage{bm}
\usepackage{subcaption}
\usepackage{graphicx}
\usepackage{svg}
\usepackage{rotating}
\usepackage{lscape}
\usepackage{array,multirow}
\usepackage[title]{appendix}
\usepackage{float}
\usepackage{stmaryrd}
\usepackage{textcomp}
\usepackage{algorithm}
\usepackage{algpseudocode}
\usepackage{amsthm}
\usepackage{listings}
\usepackage[title]{appendix}
\usepackage{import}
\usepackage{tikz}
\usetikzlibrary{quotes,arrows.meta}
\usetikzlibrary{positioning}

\usepackage{Ball}
\usepackage{Box}
\usepackage{RightBandedBox}

\usepackage{makecell}

\usepackage{xcolor,colortbl}
\usepackage{booktabs}

\usetikzlibrary{positioning}
\usetikzlibrary{3d} 

\def\checkmark{\tikz\fill[scale=0.4](0,.35) -- (.25,0) -- (1,.7) -- (.25,.15) -- cycle;}

\definecolor{codegreen}{rgb}{0,0.6,0}
\definecolor{codegray}{rgb}{0.5,0.5,0.5}
\definecolor{codepurple}{rgb}{0.58,0,0.82}
\definecolor{backcolour}{rgb}{0.95,0.95,0.92}

\lstdefinestyle{mystyle}{
  backgroundcolor=\color{backcolour},   commentstyle=\color{codegreen},
  keywordstyle=\color{magenta},
  numberstyle=\tiny\color{codegray},
  stringstyle=\color{codepurple},
  basicstyle=\ttfamily\footnotesize,
  breakatwhitespace=false,
  breaklines=true,
  captionpos=b,
  keepspaces=true,
  numbersep=0.1pt,
  showspaces=false,
  showstringspaces=false,
  showtabs=false,
  tabsize=2
}

\DeclareMathOperator{\tr}{tr}

\newtheorem{remark}{Remark}

\newcommand{\beginapp}{%
 \setcounter{equation}{0}
        \renewcommand{\theequation}{\thesection.\arabic{equation}}%
        \setcounter{table}{0}
        \renewcommand{\thetable}{\thesection.\arabic{table}}%
        \setcounter{figure}{0}
        \renewcommand{\thefigure}{\thesection.\arabic{figure}}%
        
     }
\AtBeginDocument{%
  \counterwithin{lstlisting}{section}
}
\theoremstyle{definition}

\usepackage{siunitx}
\usepackage{symbols}
\usepackage{pifont}

\usepackage{lineno}
\usepackage{natbib}

\setcitestyle{authoryear,open={(},close={)}}

\makeatletter
\def\BState{\State\hskip-\ALG@thistlm}
\makeatother

\journal{Elsevier}

\begin{document}

\begin{frontmatter}

\title{Continuous conditional generative adversarial networks for data-driven solutions of poroelasticity with heterogeneous material properties}

\author[label1,label2]{T. Kadeethum} 
\address[label1]{Sandia National Laboratories, 
   New Mexico, USA}
\address[label2]{Sibley School of Mechanical and Aerospace Engineering, Cornell University, New York, USA}

\ead{tkadeet@sandia.gov}

\author[label3]{D. O'Malley}
\address[label3]{Los Alamos National Laboratory, 
   New Mexico, USA}
\ead{omalled@lanl.gov}

\author[label4]{Y. Choi}
\address[label4]{Lawrence Livermore National Laboratory, 
   California, USA}
\ead{choi15@llnl.gov}

\author[label3]{H. S. Viswanathan}
\ead{viswana@lanl.gov}

\author[label2,label6]{N. Bouklas \corref{cor1}}
\cortext[cor1]{corresponding author}
\address[label6]{Center for Applied Mathematics,
  Cornell University,
   New York, USA}
\ead{nb589@cornell.edu}

\author[label1]{H. Yoon \corref{cor1}}
\ead{hyoon@sandia.gov}

\begin{abstract}
Machine learning-based data-driven modeling can allow computationally efficient time-dependent solutions of PDEs, such as those that describe subsurface multiphysical problems. In this work, our previous approach \citep{kadeethum2021framework} of conditional generative adversarial networks (cGAN) developed for the solution of steady-state problems involving highly heterogeneous material properties is extended to time-dependent problems by adopting the concept of continuous cGAN (CcGAN). The CcGAN that can condition continuous variables is developed to incorporate the time domain through either element-wise addition or conditional batch normalization. We note that this approach can accommodate other continuous variables (e.g., Young's modulus) similar to the time domain, which makes this framework highly flexible and extendable. Moreover, this framework can handle training data that contain different timestamps and then predict timestamps that do not exist in the training data. As a numerical example, the transient response of the coupled poroelastic process is studied in two different permeability fields: Zinn \& Harvey transformation and a bimodal transformation. The proposed CcGAN uses heterogeneous permeability fields as input parameters while pressure and displacement fields over time are model output. Our results show that the model provides sufficient accuracy with computational speed-up. This robust framework will enable us to perform real-time reservoir management and robust uncertainty quantification in realistic problems.
\end{abstract}

\begin{keyword}
non-intrusive \sep
data-driven \sep
reduced order modeling \sep
generative adversarial networks \sep
finite element \sep
poroelasticity
\end{keyword}

\end{frontmatter}


\section{Introduction}

Many subsurface applications in porous media ranging from groundwater and contaminant hydrology to $\mathrm{CO_2}$ sequestration rely on physical models
\citep{middleton2015shale,choo2018cracking,yu2020poroelastic,kadeethum2019investigation,bouklas2012swelling,newell2017investigation,kadeethum2021locally}. These models often seek the solution of the governing partial differential equations (PDEs) in a domain of interest. For instance, coupled hydro-mechanical (HM) processes in porous media, in which fluid flow and solid deformation tightly interact in a time-dependent fashion, could be described by Biot's formulation of linear poroelasticity \citep{biot1941general}. These PDEs are often solved numerically using various techniques such as finite volume or finite element methods \citep{wheeler2014coupling,deng2017locally,liu2018lowest}, which is referred to as full order model (FOM) approaches. However, computational methods to handle field-scale systems require substantial computational resources, especially when discontinuities or nonlinearities arise in the solution 
\citep{hansen2010discrete,hesthaven2016certified}. Therefore, in some instances, the FOM is not directly suitable to handle large-scale inverse problems, optimization, or even concurrent multiscale calculations in which an extensive set of simulations are required to be explored \citep{ballarin2019pod,hesthaven2016certified}. \par


A reduced order model (ROM) that aims to produce a low-dimensional representation of FOM could be an alternative to handling field-scale inverse problems, optimization, or real-time reservoir management \citep{schilders2008model,amsallem2015design,choi2019accelerating,choi2020gradient, mcbane2021component,yoon2009environmental,yoon2009numerical}. The ROM methodology primarily relies on the parameterization of a problem (i.e., repeated evaluations of a problem depending on parameters), which could correspond to physical properties, geometric characteristics, or boundary conditions \citep{ballarin2019pod,venturi2019weighted,hesthaven2016certified,hoang2021domain,copeland2021reduced,choi2020sns,choi2021space,carlberg2018conservative}. However, it is difficult to parameterize heterogeneous spatial fields of PDE coefficients such as heterogeneous material properties by a few parameters. Coupled processes such as HM processes commonly involve complex subsurface structures \citep{flemisch2018benchmarks,jia2017comprehensive, chang2020hydromechanical,chang2020operational,chang2021mitigating} where the corresponding spatially distributed material properties can span several orders of magnitude and include discontinuous features (e.g., fractures, vugs, or channels). Consequently, traditional projection-based ROM approaches might not be suitable for this type of problem as they commonly employ a proper orthogonal decomposition (POD) approach, which in turn will require a high dimensional reduced basis to capture most of the information at the expense of the computational cost \citep{kadeethum2021framework}. \par

Deep learning (DL), in particular, neural network-based supervised learning approaches, have been recently investigated for subsurface flow and transport problems \citep{zhu2018bayesian, mo2019deep, wen2021ccsnet, wen2021u, xu2021solution,kadeethum2021nonTH,wei2021aliased}. These DL approaches train various DL algorithms using training data generated by FOMs to map heterogeneous PDEs coefficients (i.e., heterogeneous permeability and/or porosity fields) and injection scenarios (e.g., injection rates and a number of wells) into state variables such as pressure and CO$_{2}$ saturation. During the online phase (i.e., prediction), these trained models are used to predict state variables, evaluate uncertainty quantification as fast forward models, or estimate material properties as inverse models \citep{kadeethum2021framework}. In most reservoir problems, these DL models can also account for time-dependent PDEs, but the output of trained models is limited to the same time interval as in the input of training data and mostly flow and transport problems. The incorporation of physical constraints (e.g., equations, relationships, and known properties) into the learning process is actively studied to improve the accuracy and training efficiency of data-driven modeling.

\cite{kadeethum2021framework} presented a data-driven generative adversarial networks (GAN) framework that can parameterize heterogeneous PDEs coefficients (i.e., heterogeneous fields), which has been demonstrated with steady-state cases for both forward and inverse problems. This GAN-based work could be considered as an extension of regression in subsurface physics through GAN model such as \cite{zhong2019predicting, laloy2019gradient, lopez2021deep}, in which heterogeneous fields are also parameterized through GAN model \citep{chan2020parametrization, hao2022siamese, guan2021reconstructing} and subsequently used to predict the state variables (pressure and displacement responses). In \cite{kadeethum2021framework}, the conditional GAN (cGAN) approach \citep{mirza2014conditional,isola2017image} was extended to the heterogeneous field for both generator and critic (i.e., discriminator) where usage of Earth mover's distance through Wasserstein loss (W loss) and gradient penalty \citep{arjovsky2017wasserstein,gulrajani2017improved} improved the model accuracy and training stability compared to the traditional GAN approach. This improvement contributed to the Earth mover's distance enforcing the cGAN framework to approximate the training data distribution rather than a point-to-point mapping. However, the framework developed by \cite{kadeethum2021framework} is limited to only steady-state solutions of given PDEs.

\begin{figure}[!ht]
  \centering
    \includegraphics[width=13.5cm,keepaspectratio]{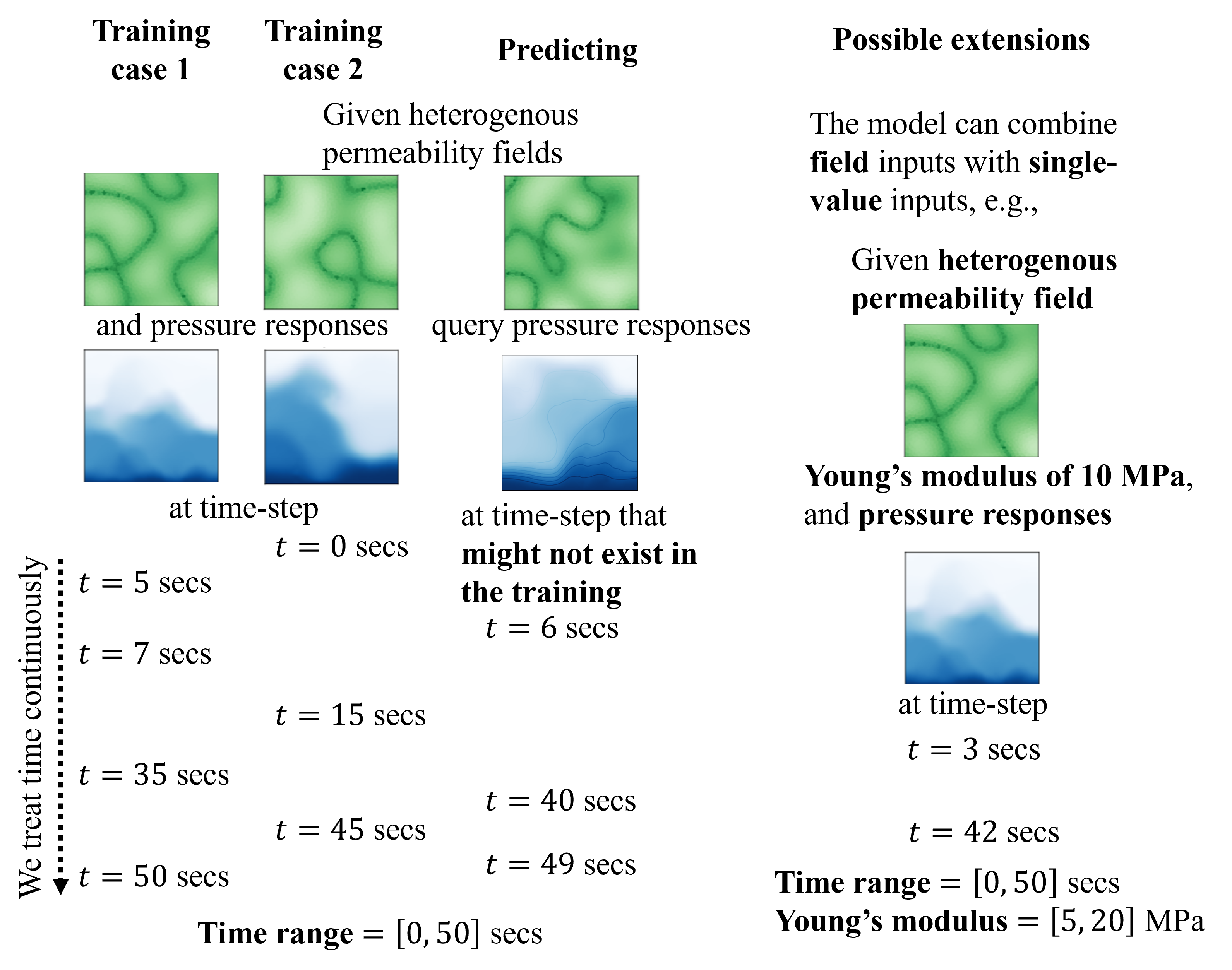} 
  \caption{The main characteristics of this model. Our model treats the time domain as a continuous variable, which means our training data could have different time-step. Moreover, during the prediction, our model can predict responses at time-step that does not exist in the training data. This model could be extended to include any continuous variables (e.g., Young's modulus) the same way it treats the time domain.}
  \label{fig:novelty}
\end{figure}

Recently, \cite{ding2020continuous} developed continuous cGAN (CcGAN) to condition the GAN model with continuous variables such as quantitative measures (e.g., the weight of each animal) rather than  categorical data (e.g., cat or dog). For PDE problems, the concept of CcGAN can also be extended to quantify continuous variables (e.g., time domain), enabling the solution of time-dependent PDEs. In this work, we extend our previous work \citep{kadeethum2021framework} by extending the CcGAN concept to solve for time-dependent PDEs. This new framework is developed by utilizing element-wise addition or conditional batch normalization \citep{de2017modulating} to incorporate the time domain in both training and prediction processes. 
As presented in Figure \ref{fig:novelty}, our model treats the time domain as a continuous variable. Therefore, this model can handle the training data that contains different time-step resolutions. Furthermore, we can predict the continuous-time response without being limited to time instances that correspond to the training data. This design also provides flexibility to incorporate other parameterized continuous variables (e.g., Young's modulus, boundary conditions) as parameters to our framework. 

The CcGAN approach to solving time-dependent PDEs in this work is uniquely employed to develop a data-driven surrogate model given highly heterogeneous permeability fields generated from two known distributions. The CcGAN performance will be evaluated by comparing the CcGAN-based results with FOM-based solutions, highlighting the computational accuracy and efficiency in heterogeneous permeability fields. The rest of the manuscript is summarized as follows. We begin with the model description and CcGAN architecture in Section \ref{sec:method}. The two variants of the framework, including element-wise addition or conditional batch normalization to incorporate the time domain, are also discussed. We then illustrate our ROM framework using three numerical examples in Section \ref{sec:numer_results}. Lastly, we conclude our findings in Section \ref{sec:conclusions}.

\section{Methodology}\label{sec:method}

\subsection{General governing equations}

We first present a general framework with a parameterized system of time-dependent PDEs and then as a demonstration case we focus on the linear poroelasticity to represent a coupled solid deformation and fluid diffusion in porous media with highly heterogeneous permeability fields. A parameterized system of time-dependent PDEs are as following

\begin{equation} \label{eq:gen_pdes}
\begin{split}
\bm{F}\left( t^n, \bm{\mu}^{(i)}\right) = \bm{0}  &\text { \: in \: } \Omega, \\
\bm{X} =\bm{f}_{D} &\text { \: on \: } \partial \Omega_{D},\\
- \nabla \bm{X} \cdot \mathbf{n}=\bm{f}_N &\text { \: on \: } \partial \Omega_{N}. \\
\bm{X}=\bm{X}_{0} &\text { \: in \: } \Omega \text { at } t^n = 0,
\end{split}
\end{equation}

\noindent
where $\bm{F}\left( \cdot \right)$ corresponds to the system of time dependent PDEs, $\Omega \subset \mathbb{R}^{n_d}$ (${n_d} \in \{1,2,3\}$) denotes the computational domain, $\partial \Omega_{D}$ and $\partial \Omega_{N}$ denote the Dirichlet and Neumann boundaries, respectively. $\bm{f}_{D}$ and $\bm{f}_N$ are prescribed values on $\partial \Omega_{D}$ and $\partial \Omega_{N}$, respectively. $\bm{X}_{0}$ is an initial value of $\bm{X}$. The time domain $\mathbb{T} = \left(0, \tau\right]$ is partitioned into $N^t$ subintervals such that $0=: t^{0}<t^{1}<\cdots<t^{N} := \tau$, We denote $t^{n} \in \mathbb{T}$ as $n$th time-step, $n\in[0,N]$. $\bm{X}$ is a set of scalar ($\bm{X} \in \mathbb{R}$) or tensor valued (e.g. $\bm{X} \in 
\mathbb{R}^{n_d}\,\,\mathrm{or}\,\,\mathbb{R}^{n_d}\times\mathbb{R}^{n_d}$) generalized primary variables. For the parameter domain $\mathbb{P}$, it composes of $\mathrm{M}$ members, i.e., $\bm{\mu}^{(1)}$, $\bm{\mu}^{(2)}$, $\dots$, $\bm{\mu}^{(\mathrm{M-1})}$, $\bm{\mu}^{(\mathrm{M})}$, and $\bm{\mu}^{(i)}$ could be any instances of $\bm{\mu}$ given $i = 1, \dots, \mathrm{M}$. $\bm{\mu}^{(i)}$ could be scalar ($\bm{\mu}^{(i)} \in  \mathbb{R}$) or tensor valued (e.g. $\bm{\mu}^{(i)} \in 
\mathbb{R}^{n_d}\,\,\mathrm{or}\,\,\mathbb{R}^{n_d}\times\mathbb{R}^{n_d}$) generalized parameters. We want to emphasize that $\bm{X}$ is an exact solution of $\bm{F}\left( \bm{X}; t^n, \bm{\mu}^{(i)}\right)$ and ${\bm{X}_h}$ is an approximation obtained from FOM. In general, $\bm{\mu}^{(i)}$ could correspond to physical properties, geometric characteristics, or boundary conditions at any given time $t$. In this study, we limit our interest in approximate primary variables $\bm{X}_h$ for a solution of the system of PDEs given the generalized parameters $\bm{\mu}$ such as heterogeneous permeability fields that are constant through time $t$. \par

\subsection{Framework development}

\begin{figure}[!ht]
  \centering
    \includegraphics[width=13.5cm,keepaspectratio]{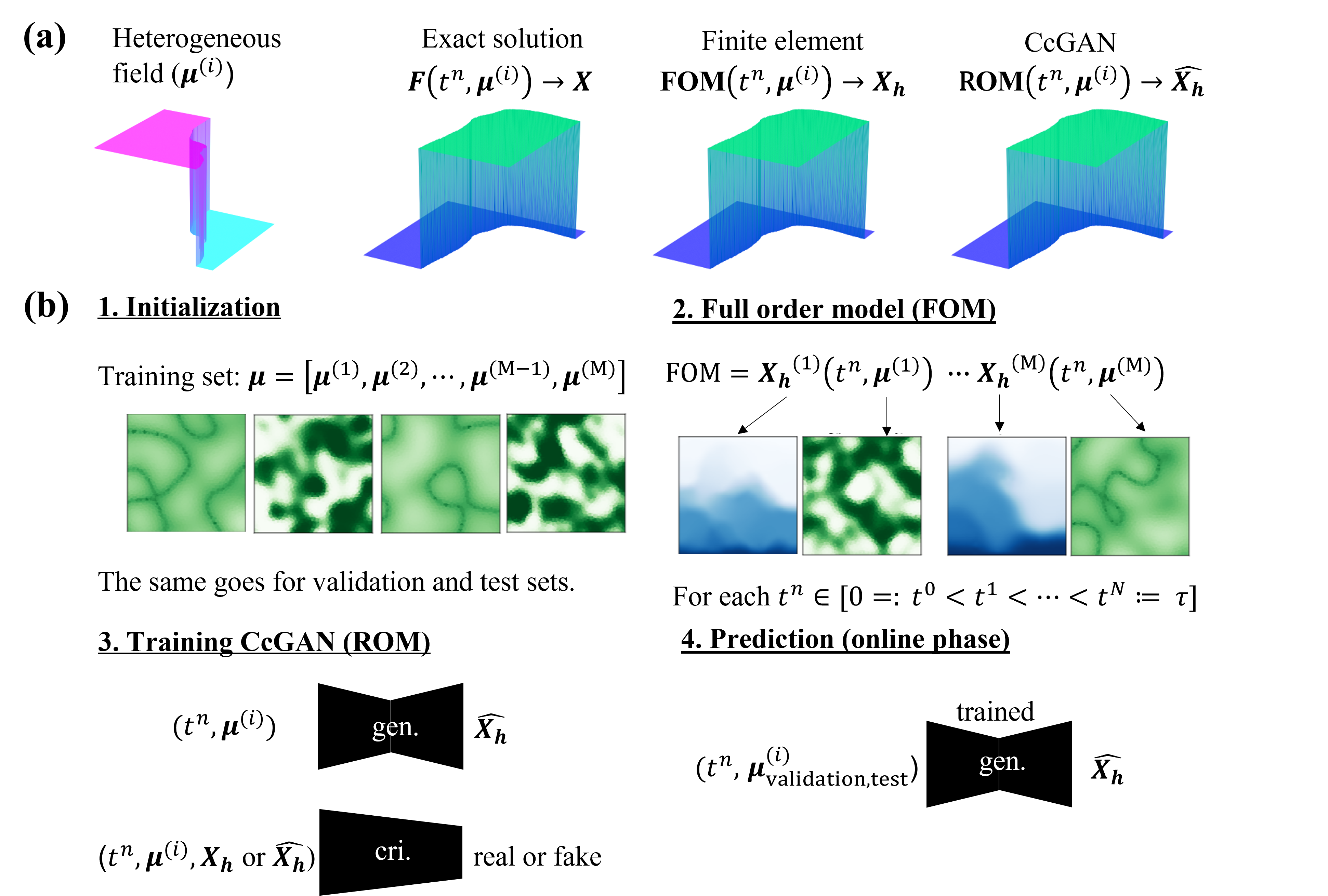} 
  \caption{The main idea and procedures taken in developing the framework is shown. Gen. represents generator, and cri. is critic. In $\bm{\mathrm{(a)}}$, $\bm{X}$ is an exact solution with given $t^n$ and $\bm{\mu}^{(i)}$, $\bm{X}_h$ is an approximation of $\bm{X}$ by FOM, and $\widehat{\bm{X}_h}$ is an approximation of $\bm{X}$. $\bm{\mu}^{(i)}$ here represents a heterogeneous permeability field. Our goal is to develop a proxy that could provide $\widehat{\bm{X}_h}$ that is as close as possible to $\bm{X}_h$, but requires a much cheaper computational cost. In $\bm{\mathrm{(b)}}$, we illustrate procedures taken to develop a data-driven solution of time-dependent coupled hydro-mechanical processes using continuous conditional generative adversarial networks. }
  \label{fig:intro}
\end{figure}

As in a conceptual schematic (Figure \ref{fig:intro}), we train our framework using $\bm{X}_h$ obtained from FOM to deliver $\widehat{\bm{X}_h}$ with acceptable accuracy and high computational efficiency. The proposed framework consists of (1) the offline phase starting from data generation of permeability fields and $\bm{X}_h$ to  training of our proposed CcGAN and (2) the online phase of predicting $\widehat{\bm{X}_h}$ as presented in Figure \ref{fig:intro}b. \par

\subsubsection{Offline stage}

The first step is an initialization of training, validation, and test sets of parameters ($\bm{\mu}_\mathrm{training}$, $\bm{\mu}_\mathrm{validation}$, and $\bm{\mu}_\mathrm{test}$) of cardinality $\mathrm{M}_\mathrm{training}$, $\mathrm{M}_\mathrm{validation}$, and $\mathrm{M}_\mathrm{test}$, respectively. We illustrate only $\bm{\mu}$ and $\bm{\mu}_\mathrm{test}$ in Figure \ref{fig:intro} and omit a subscript of training in $\bm{\mu}_\mathrm{training}$ and $\mathrm{M}_\mathrm{training}$ hereinafter for the sake of brevity. We want to emphasize that $\bm{\mu}$ $\cap$ $\bm{\mu}_\mathrm{validation}$ $=$ $\varnothing$, $\bm{\mu}$ $\cap$ $\bm{\mu}_\mathrm{test}$ $=$ $\varnothing$, and $\bm{\mu}_\mathrm{validation}$ $\cap$ $\bm{\mu}_\mathrm{test}$ $=$ $\varnothing$. These $\bm{\mu}$, $\bm{\mu}_\mathrm{validation}$, and $\bm{\mu}_\mathrm{test}$ here represent any physical properties, but it could also serve as geometric characteristics or boundary conditions. In this work, we follow \cite{kadeethum2021framework} and focus on using $\bm{\mu}$, $\bm{\mu}_\mathrm{validation}$, and $\bm{\mu}_\mathrm{test}$ to represent collections of spatially heterogeneous scalar coefficients - more specifically heterogeneous permeability fields as described in the Section \ref{sec:data_generation} for data generation.

In the second step, we query the FOM, which can provide a solution in a finite-dimensional setting for each parameter $\bm{\mu}$ (i.e., $\bm{\mu}^{(i)}$) in the training set. Throughout this study, for the sake of simplicity, we use a uniform time-step which leads to each query of the FOM having the same number of $N^t$. However, as presented in Figure \ref{fig:novelty}, our framework could handle cases where adaptive time-stepping is required, for instance, advection-diffusion problems. The same operations follow for each $\bm{\mu}_\mathrm{validation}$ and $\bm{\mu}_\mathrm{test}$ in the validation and test sets. This work focuses on the linear poroelasticity equations and demonstrates our proposed framework with highly heterogeneous permeability fields. The FOM is used to approximate primary variables $\bm{X}_h$, which correspond to bulk displacement ($\bm{u}_h$) and fluid pressure ($p_h$) fields at each time-step $t^n$ given the field of parameters $\bm{\mu}^{(i)}$, in this case - permeability field, as input. Please find the detailed description in Appendix \ref{sec:prob_description}. \par

\begin{figure}[!ht]
  \centering
    \includegraphics[width=13.5cm,keepaspectratio]{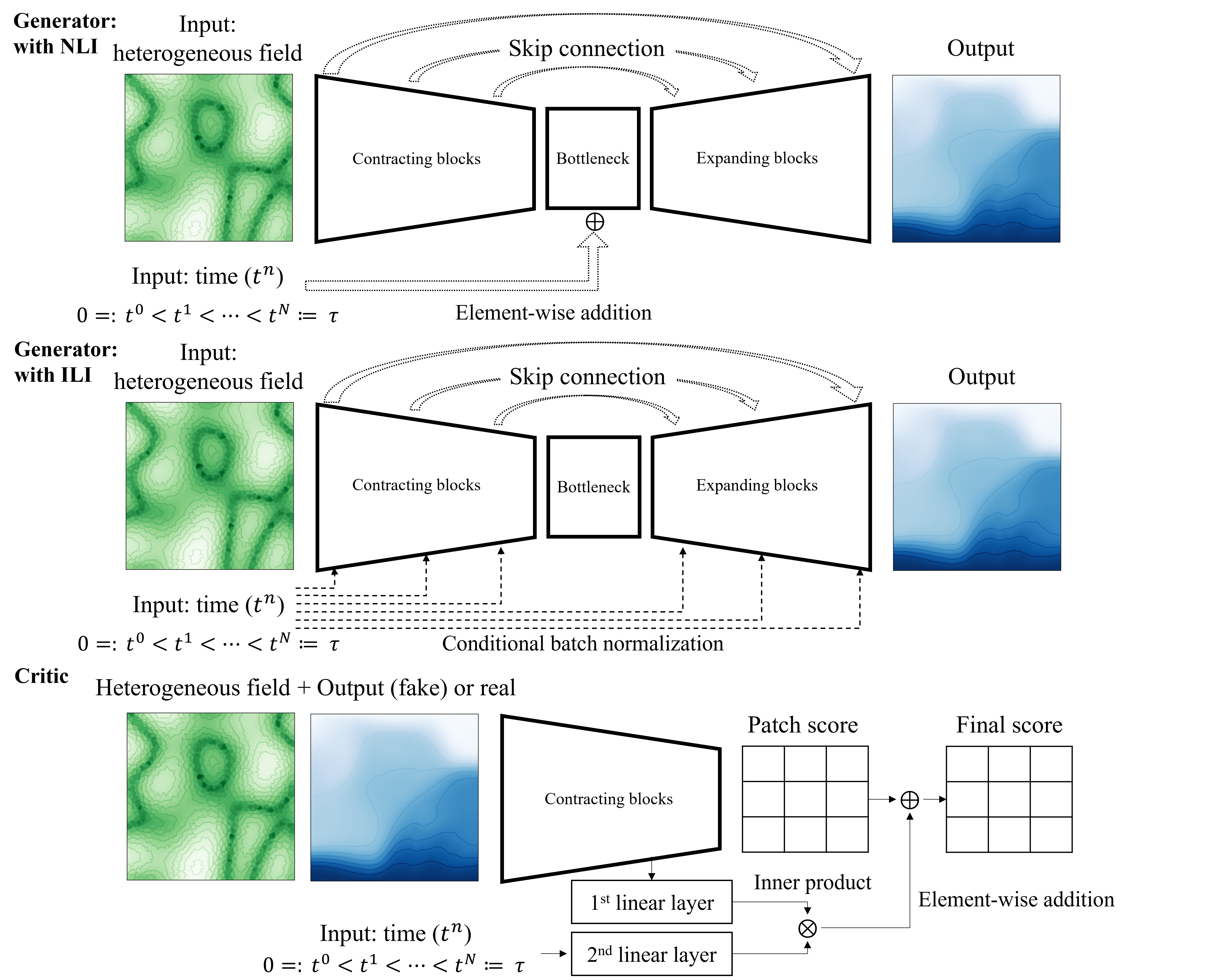} 
  \caption{ROM for time-dependent PDEs using continuous conditional generative adversarial networks (CcGAN). We note that the critic is similar for both models (generator with NLI and generator with ILI).}
  \label{fig:model}
\end{figure}

In the third step, ROM is constructed by training the data generated from the FOM where the inputs to the model are $t^n$ and $\bm{\mu}^{(i)}$, and the output is $\bm{u}_h$ or $\bm{X}_{h}$ with given $t^n$ and $\bm{\mu}^{(i)}$. In this study, we build a separate model for each primary variable ($\bm{u}_h$ and $p_h$), although both primary variables can be trained together with a single model. A key aspect of this work is to apply the CcGAN image-to-image translation framework for time-dependent PDEs by adapting the concept of a naive label input (NLI) and an improved label input (ILI) proposed by \cite{ding2020continuous} to the framework developed by \cite{kadeethum2021framework}. The proposed framework in this work consists of a \emph{generator} and \emph{critic} where two types of architecture for the \emph{generator} with a similar \emph{critic} architecture are presented (Figure \ref{fig:model}). 

The first one uses the NLI concept (i.e., NLI model) by introducing a temporal term ($t^{n} \in \mathbb{T}$) to the generator's bottleneck using element-wise addition. The details of the architecture can be found in Table \ref{tab:unet_model1} and Listing \ref{list:nli}. The second one adopts the ILI concept (i.e., ILI model) by injecting the temporal term to all layers inside the generator through conditional batch normalization \citep{de2017modulating}. However, in contrast to \cite{ding2020continuous} and \cite{de2017modulating}, our $t^{n}$ is not categorical data (i.e., not a tag of number ranging from zero to nine), but continuous variable (i.e., $t^{n} \in \mathbb{T}$). Hence, we replace embedded layers with an artificial neural network (ANN). To elaborate, each conditional batch normalization composes of one ANN and one batch normalization layer. Each ANN composes of one input, three hidden, and one output layers. Each hidden layer and the input layer are subjected to the tanh activation function. In this way, we can inform each $t^{n}$ to each layer of our generator through a conditional batch normalization concept, see Listing \ref{list:cbn} for the implementation. The details of the ILI architecture can be found in Table \ref{tab:unet_model2} and Listing \ref{list:ili}. \par

The critic, similar for both NLI and ILI, uses time $t^n$, parameter  $\bm{\mu}^{(i)}$, and primary variable ($\bm{u}_h$ or $p_h$) as its inputs. The output is a patch score added with an inner product calculated using two linear layers and output from the last contracting block ($4^{\mathrm{th}}$ contracting block) of the critic. To elaborate, parameter ($\bm{\mu}^{(i)}$) and primary variable ($\bm{u}_h$ or $p_h$) are fed into the $1^{\mathrm{st}}$ convolutional layer of the critic while time ($t^n$) is injected into the model using $2^{\mathrm{nd}}$ linear layer shown in Figure \ref{fig:model}. The output from the $4^{\mathrm{th}}$ contracting block of the critic is then passed through the $1^{\mathrm{st}}$ linear layer shown in Figure \ref{fig:model} and performed an inner product operation with the output from the $2^{\mathrm{nd}}$ linear layer. The result of this inner product is then added (element-wise) to the patch score presented in Figure \ref{fig:model}. The architecture of the critic can be found in Table \ref{tab:disc} and Listing \ref{list:critic}. \par

To train both generator and critic, we normalize $t^n$, $\bm{\mu}^{(i)}$, and primary variables ($\bm{X}_{h}$ or, in this paper, $\bm{u}_h$ and $p_h$) to be in a range of $[0, 1]$. The Wasserstein (W) loss (\cite{arjovsky2017wasserstein,gulrajani2017improved}) is used since it has been shown to provide the best result when dealing with building data-driven frameworks for PDEs as shown in \cite{kadeethum2021framework}. In short, this implementation enforces the model to approximate the training data distribution instead of aiming to do the point-to-point mapping. This improves our model generalization, which is essential as we deal with heterogeneous permeability fields. The W loss is expressed as  \par

\begin{equation} \label{eq:total_min_max_add_wloss}
\min _{G} \max _{C} \left[ \ell_{a} +\lambda_{{r}}\ell_r  +\lambda_{{p}}\wp_p \right].
\end{equation}

\noindent
Here, $G$ and $C$ are short for generator and critic, respectively, and $\ell_{a}$ is the Earth mover's distance defined as

\begin{equation} 
\ell_{a} = \frac{1}{\mathrm{B}} \sum_{i = 1}^{\mathrm{B}} C(t^i,\bm{\mu}^{(i)},{\bm{X}_{h}}_i) -\frac{1}{\mathrm{B}} \sum_{i = 1}^{\mathrm{B}} C\left(t^i,\bm{\mu}^{(i)},\widehat{{\bm{X}_{h}}_i} \right), 
\end{equation}

\noindent
where $C\left( \cdot \right)$ is the final score of the critic with given $t^i$, $\bm{\mu}^{(i)}$, and ${\bm{X}_{h}}_i$ or $\widehat{{\bm{X}_{h}}_i}$ shown in Figure \ref{fig:model}, and $\mathrm{B}$ is a batch size, which is set as $\mathrm{B}=4$. The Earth mover's distance is used to measure the distance between output of the model and the training data. This way helps our model to better generalize its prediction. Additionally, $\lambda_r$ is a user-defined penalty constant that we set at $\lambda_r=500$, and $\ell_r$ as a reconstruction error term is given by

\begin{equation} \label{eq:l1_loss}
\ell_r= \frac{1}{\mathrm{B}} \sum_{i = 1}^{\mathrm{B}}  \left|\widehat{{\bm{X}_{h}}_{i}}-{\bm{X}_{h}}_{i}\right|.
\end{equation}

\noindent
$\lambda_{{p}}$ denotes a gradient penalty constant set to 10 throughout this study; $\wp_p$ is the gradient penalty regularization. The latter is used to enforce Lipschitz continuity of the weight matrices ($\mathbf{W}$), i.e., Euclidean norm of discriminator’s gradient is at most one, and it reads as

\begin{equation}
   \wp_p = \frac{1}{\mathrm{B}} \sum_{i = 1}^{\mathrm{B}}\left(\|\nabla C({t^i,\bm{\mu}}_i, \Bar{{\bm{X}_{h}}}_i)\|_{2}-1\right)^{2},
\end{equation}

\noindent
where $\| \cdot \|_{2}$ is L2 or Euclidean norm. This term helps to improve the stability of the training by limiting the step we can take in updating our trainable parameters (weight matrices ($\mathbf{W}$) and biases ($\bm{b}$). The term $\Bar{{\bm{X}_{h}}}_i$ is an interpolation between $\widehat{{\bm{X}_{h}}}_i$ and ${\bm{X}_{h}}_i$, which is defined by

\begin{equation}
   \Bar{{\bm{X}_{h}}} = {\epsilon}_i {\bm{X}_{h}} + (1 -  {\epsilon}_i)\widehat{{\bm{X}_{h}}}_i.
\end{equation}

\noindent
We randomly select $\epsilon_i$ for each $\Bar{{p_h}}_i$ from a uniform distribution on the interval of $[0, 1)$. We use the adaptive moment estimation (ADAM) algorithm \citep{kingma2014adam} to train  the framework (i.e., updating a set of weight matrices ($\mathbf{W}$) and biases ($\bm{b}$)). The learning rate ($\eta$) is calculated as \cite{loshchilov2016sgdr}

\begin{equation}
\eta_{c}=\eta_{\min }+\frac{1}{2}\left(\eta_{\max }-\eta_{\min }\right)\left(1+\cos \left(\frac{\mathrm{step_c}}{\mathrm{step_f}} \pi\right)\right)
\end{equation}

\noindent
where $\eta_{c}$ is a learning rate at $\mathrm{step_c}$, $\eta_{\min }$ is the minimum learning rate ($1 \times 10^{-16}$), $\eta_{\max }$ is the maximum or initial learning rate ($1 \times 10^{-4}$), $\mathrm{step_c}$ is the current step, and $\mathrm{step_f}$ is the final step. We note that each step refers to each time we perform back-propagation, including updating both generator and critic's parameters ($\mathbf{W}$ and $\bm{b}$).    \par

\subsubsection{Online stage}

For the fourth step, we use the \emph{trained} generator to predict $\widehat{\bm{X}_{h}}$ given $t^n$ and $\bm{\mu}_\mathrm{validation}$ or $\bm{\mu}_\mathrm{test}$ for the instances of the parameters belonging to the validation and test sets. To avoid over-fitting, we first use $\bm{\mu}_\mathrm{validation}$ to evaluate our framework as a function of epoch. Subsequently, we select the model (fixed $\mathbf{W}$ and $\bm{b}$ at a certain epoch) that provides the best accuracy for $\bm{\mu}_\mathrm{validation}$ to test the $\bm{\mu}_\mathrm{test}$. To elaborate, we train our model for 50 epochs. We then test our model against our validation set and observe the model performance as a function of the epoch. We then select a set of $\mathbf{W}$ and $\bm{b}$ at the epoch that has the best accuracy. Other hyper-parameters including a number of convoluational neural netwrok (CNN) layers, a number of hidden layers, and CNN parameters, and initialization of the framework are used based on the study in \cite{kadeethum2021framework}. \par

\begin{remark} \label{remark:1}
As presented in Figure \ref{fig:novelty}, by treating the time domain as a continuous variable, our framework could be trained using training data that contains different time-step. Furthermore, during our online inquiry, we simply interpolate at the time of interest within the time domain provided during the training phase, which may or may not exist in the training data. This characteristic is an asset of our framework because our framework is not bound by a time-stepping scheme that traditional numerical analysis or other data-driven framework has \citep{zhu2018bayesian, mo2019deep, wen2021ccsnet, xu2021solution}. Our framework can evaluate quantities of interest at any time required. For instance, we may be interested in a pressure field at one, two, and three hours with a given permeability field. To achieve that using FOM, one may need to go through many \textit{intermediate} steps in between to satisfy, but our framework could evaluate any particular time-step immediately within training data we evaluate.
\end{remark}

\section{Data generation}\label{sec:data_generation}

We utilize a discontinuous Galerkin finite element (FE) model of linear poroelasticity developed in \cite{kadeethum2020enriched, kadeethum2020finite} to generate training, validation, and test sets, see Figure \ref{fig:intro} - initialization. The geometry, boundary conditions, and input parameters are similar to that used in \cite{kadeethum2021framework} where a steady-state solution of linear poroelasticity is studied, but, in this work, the temporal output of pressure and displacement are investigated, resulting in the dynamic behavior of pressure ($p_h$), displacement ($\bm{u}_h$) as well as pore volume. The mesh and boundary conditions over the square domain used in this work are presented in Figure \ref{fig:mesh}. We enforce constant pressures of 0 and 1000 \si{Pa} at the top and bottom boundaries, respectively, to allow fluid to flow from the bottom to the top while no-flow boundary on both left and right sides. Furthermore, we compress the medium with normal traction of 1000 \si{Pa} applied at the top boundary. We fix the normal displacement to zero \si{m} for the left, right, and bottom boundaries. The initial pressure is 1000 \si{Pa}, and initial displacement is calculated based on the equilibrium state. \par

\begin{figure}[!ht]
   \centering
         \includegraphics[keepaspectratio, height=11.2cm]{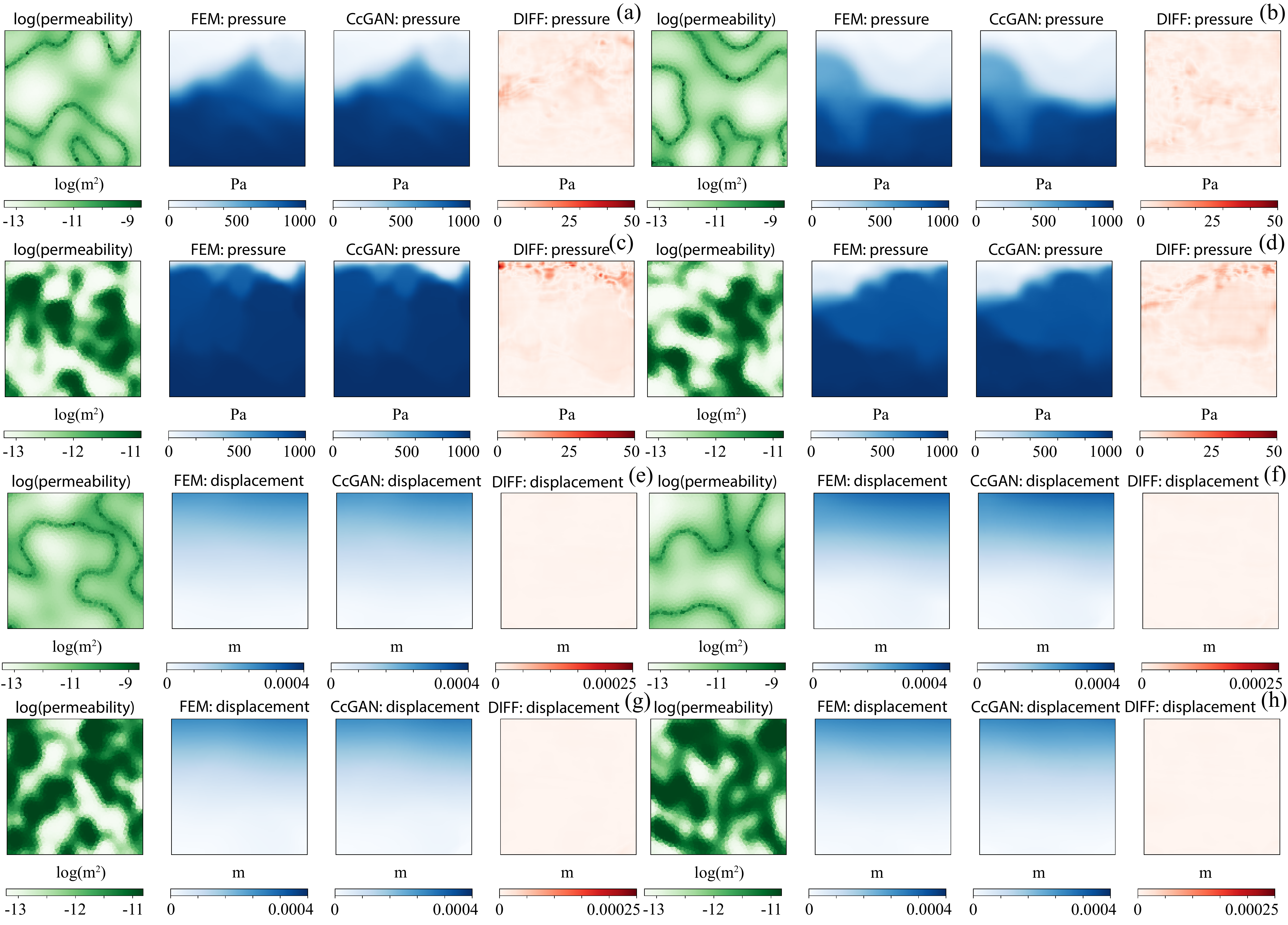}
  \caption{Comparison of pressure (a-d) and displacement (e-h) results between the FEM (FOM) and the CcGAN (ROM) given the permeability field. For pressure results, (a-b) Zinn \& Harvey transformation cases at $t=50$ and $t=225$, (c-d) bimodal transformation cases at $t=25$ and $t=250$. For displacement results, (e-f) Zinn \& Harvey transformation cases at $t=50$ and $t=225$, (g-h) bimodal transformation cases at $t=25$ and $t=250$. Each example is randomly selected at different times given the permeability field from 1000 test sets to show that our model can evaluate at any time. The results are shown from the test set of the ILI framework trained with $\mathrm{M}$ = 20,000 examples (Example 3) where each Zinn \& Harvey (Example 1) and bimodal (Example 2) transformation has 10,000 examples as presented in Table 1. The fluid flows from the bottom to the top surface. The media is compressed from the top. The rest of the surfaces, left, right, and bottom, are allowed to be moved only in the normal direction. More details can be found in Appendix A. Note that the best model used for the test set is selected based on the performance of the validation set and the permeability ranges of each transformation are different.}
   \label{fig:pics_of_test_com}
\end{figure}

To obtain a set of parameters $\bm{\mu}$ corresponding to heterogeneous $\bm{k}$ fields, we focus on two types of highly heterogeneous $\bm{k}$ fields generated using: (1) a Zinn \& Harvey transformation \citep{zinn2003good}, and (2) a a bimodal transformation \citep{muller2020}. The details of generation of $\bm{k}$ fields is available in \cite{kadeethum2021framework}. Briefly, the $\bm{k}$ field from the Zinn \& Harvey transformation has a wider range of $\bm{k}$ values with thinner high permeability pathways. This feature represents highly heterogeneous sedimentary aquifers with preferential flow pathways, such as the MADE site in Mississippi \citep{zinn2003good} and the Culebra dolomite developed for the Waste Isolation Pilot Plant (WIPP) project in New Mexico \citep{yoon2013parameter}. In contrast, the $\bm{k}$ field from the bimodal transformation has narrow range $\bm{k}$ values with wider high permeability pathways, which is a good representation of sandstone reservoirs with an iron inclusion, for example, Chadormalu reservoirs in Yazd province, Iran \citep{daya2015application}. A few examples of $\bm{k}$ fields from both transformations are shown in Figures \ref{fig:pics_of_test_com}. \par

In this work, three examples of $\bm{k}$ fields are used. Two examples are from Zinn \& Harvey (Example 1) and bimodal (Example 2) distributions. For Example 3, these two $\bm{k}$ fields are used together. Note that we employ unstructured grids in the finite element solver. However, our framework in this study requires a structured data set. Thus, we interpolate the FE result $p_h$ to structured grids using cubic spline interpolation. We then replace the FOM dimension ${N}_h^p$, associated with the unstructured grid, with a pair $(\widetilde{N}_h^p, \widetilde{N}_h^p) = (128, 128)$, corresponding to the structured grid. The same procedures are carried out for the displacement field $\bm{u}_h$. \par

For simplicity, the FE solver employs a fixed number of $N^t$ for all $\bm{k}^{(i)} \in \bm{k}$. Our time domain is set as $\mathbb{T} = \left(0, 250 \right]$ \si{seconds}, and $N^t = 10$, which leads to $\Delta t = 25$ \si{seconds}. 
The total size of data set is calculated by $\mathrm{M}_\mathrm{i} N^t$ where $\mathrm{i}$ is the number of training, validation, or test sets. While we investigate the effect of $\mathrm{M}$ on the data-driven model accuracy, $\mathrm{M}_\mathrm{validation} = \mathrm{M}_\mathrm{test} = 500$ is fixed. The samples of the test set of $\bm{k}^{(i)}$, $p_h\left( t^n, \bm{k}^{(i)} 
\right)$, and $\bm{u}_h\left( t^n, \bm{k}^{(i)}  \right)$ are presented in Figure \ref{fig:pics_of_test_com}. We note that the difference (DIFF) between solutions produced by the FOM (FEM in this case) and ROM (CcGAN in this case) is calculated by

\begin{equation}\label{eq:diff}
\operatorname{DIFF}_{\bm{X}}(t^n, \bm{\mu}_\mathrm{test}^{(i)})= \left|\bm{X}_h(:, t^n, \bm{\mu}_\mathrm{test}^{(i)}) - \widehat{\bm{X}}_h(:, t^n, \bm{\mu}_\mathrm{test}^{(i)})\right|.
\end{equation}

\noindent
To reiterate, $\bm{X}_h$ represents $p_h$ and $\bm{u}_h$, $\widehat{\bm{X}}_h$ represents $\hat{p}_h$ and $\hat{\bm{u}}_h$, and $\bm{\mu}_\mathrm{test}^{(i)}$ is $\bm{k}_\mathrm{test}^{(i)}$ field. We also use the relative root mean square error (relative RMSE) between $x_{i}$ (FOM) and $\hat{x}_{i}$ (ROM) to evaluate the model performance as

\begin{equation}\label{eq:rmse}
\mathrm{relative \: \: RMSE}=\frac{\sqrt{\frac{\sum_{i=1}^{\mathrm{M}}\left(x_{i}-\hat{x}_{i}\right)^{2}}{\mathrm{M}}}}{{\sqrt{\frac{\sum_{i=1}^{\mathrm{M}} x_{i}^2} {\mathrm{M}}}}}, \quad x_i \in \bm{X}_h \: \mathrm{and} \: \hat{x}_{i} \in \widehat{\bm{X}_h}.
\end{equation}

\section{Results and discussion}\label{sec:numer_results}

\subsection{Example 1: Zinn \& Harvey transformation}

\begin{figure}[!ht]
   \centering
         \includegraphics[keepaspectratio, height=4.0cm]{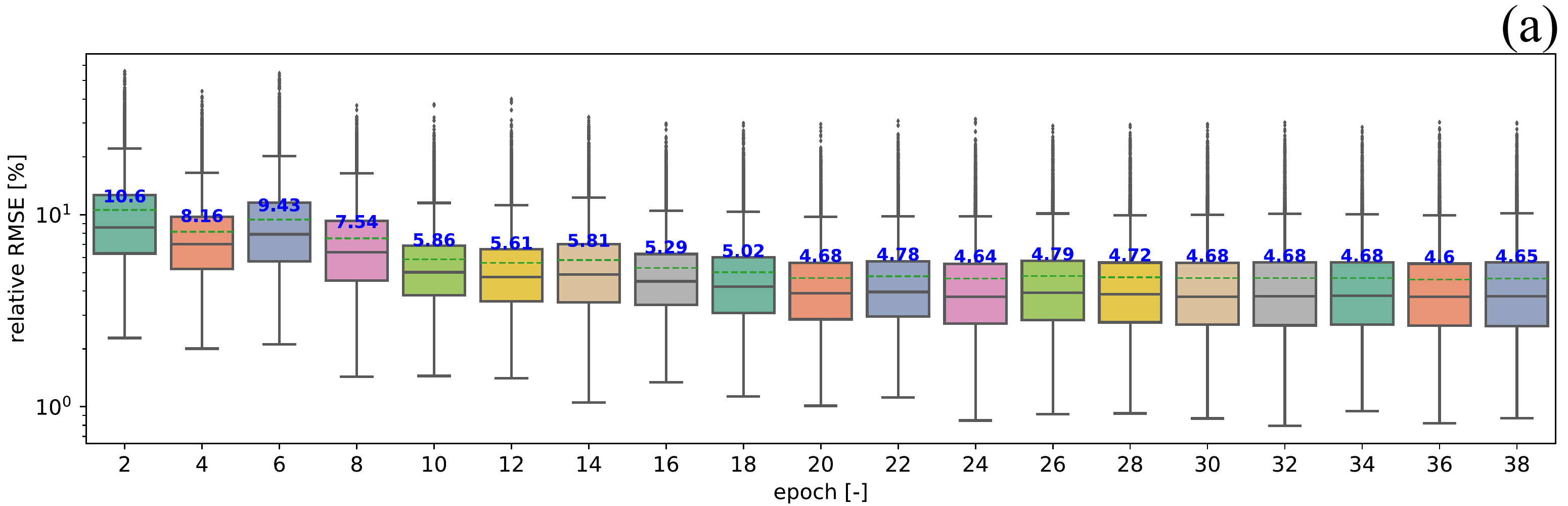}
         \includegraphics[keepaspectratio, height=4.0cm]{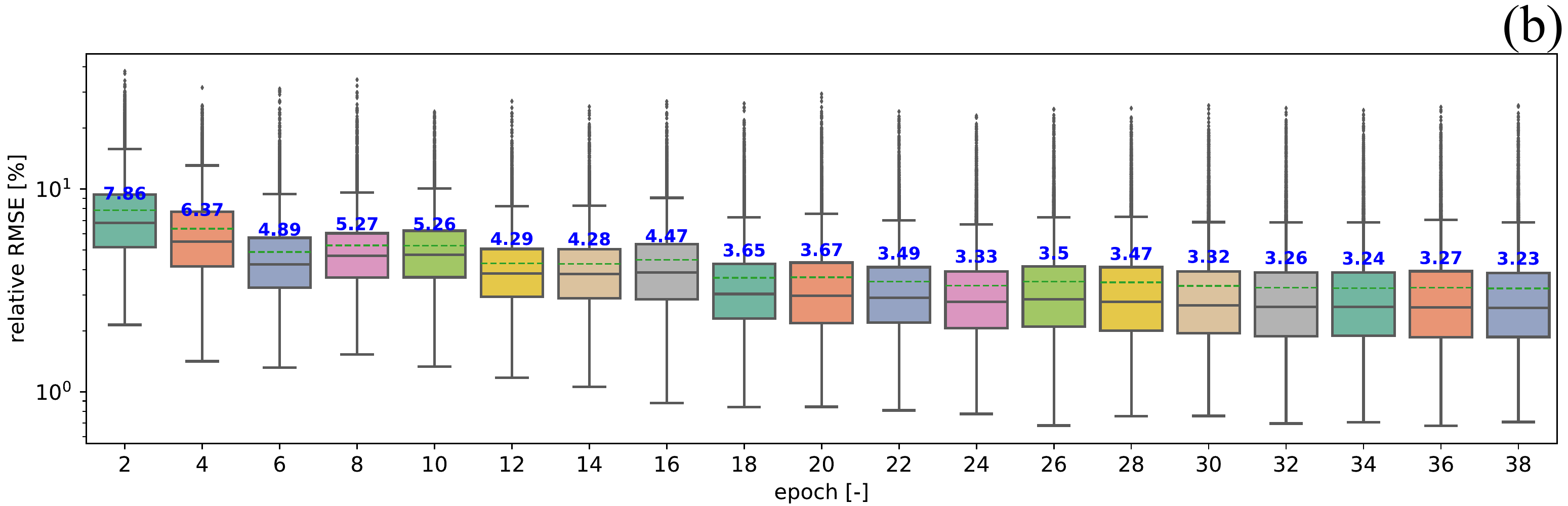}
         \includegraphics[keepaspectratio, height=4.0cm]{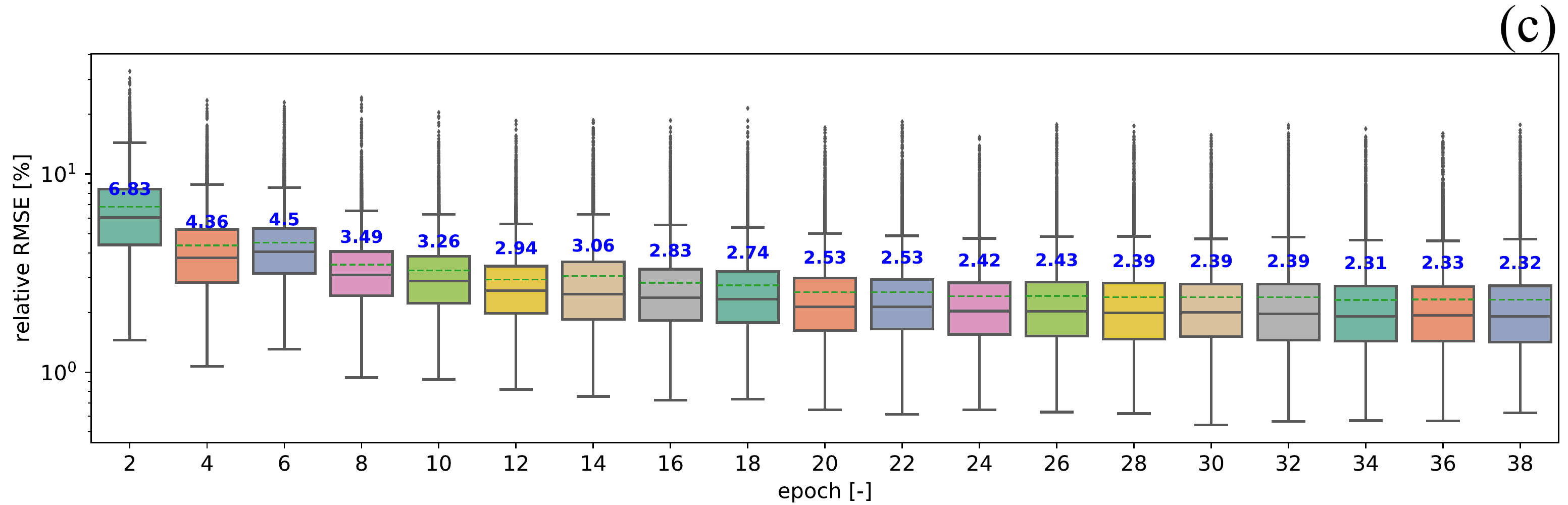}
         \includegraphics[keepaspectratio, height=4.0cm]{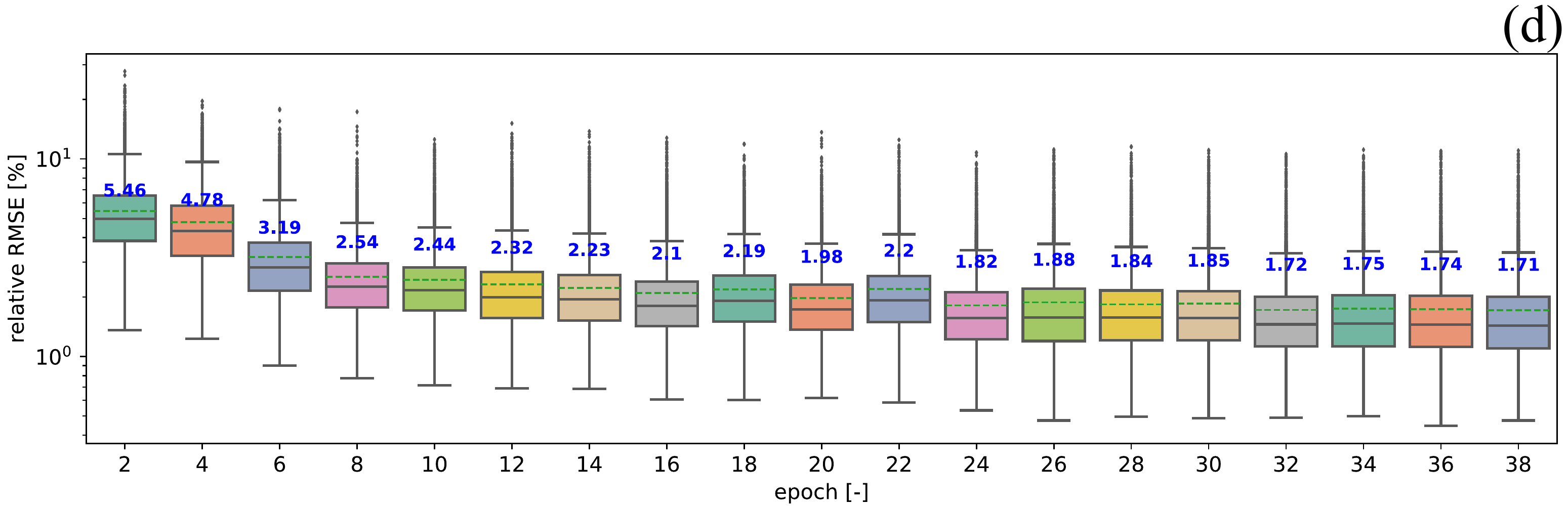}
   \caption{Example 1: relative Root Mean Square Error (relative RMSE) of NLI for (a) $\mathrm{M} = 1250$, (b) $\mathrm{M} = 2500$, (c) $\mathrm{M} = 5000$, and (d) $\mathrm{M} = 10000$. These results are calculated based on the validation set (500 samples). We note that the black dots represent outliers, and the box plot covers the interval from the 25th percentile to 75th percentile, highlighting the mean (50th percentile) with an black line. Blue line and blue text represent a mean value.}
   \label{fig:ex1_val_model1}
\end{figure}

\begin{figure}[!ht]
   \centering
         \includegraphics[keepaspectratio, height=4.0cm]{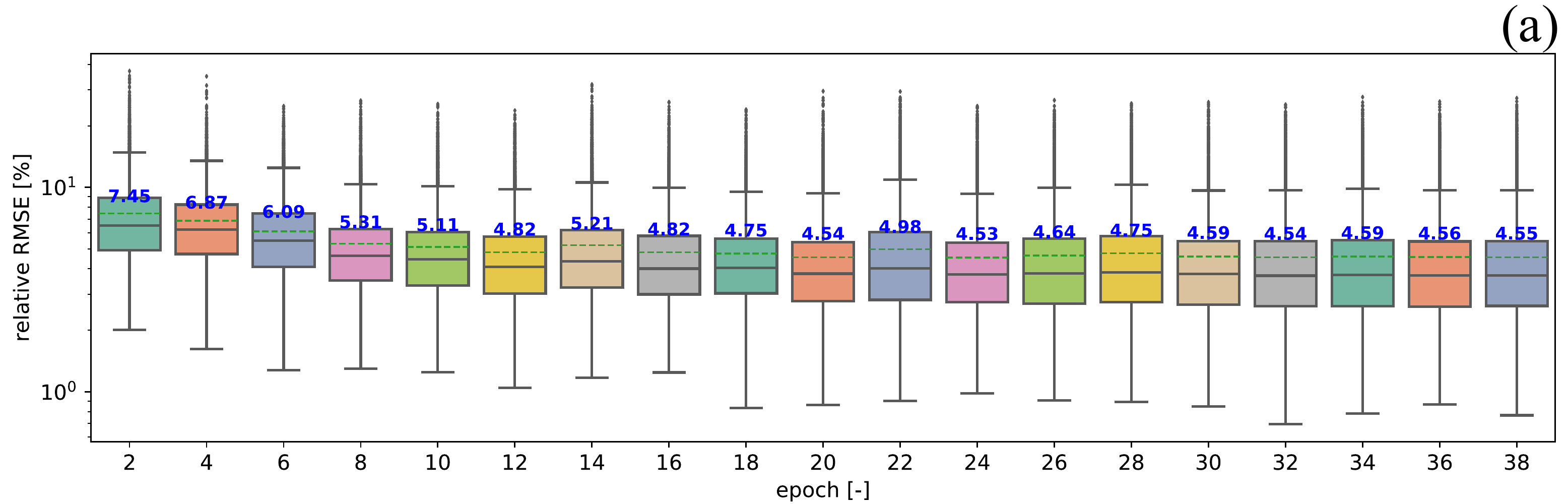}
         \includegraphics[keepaspectratio, height=4.0cm]{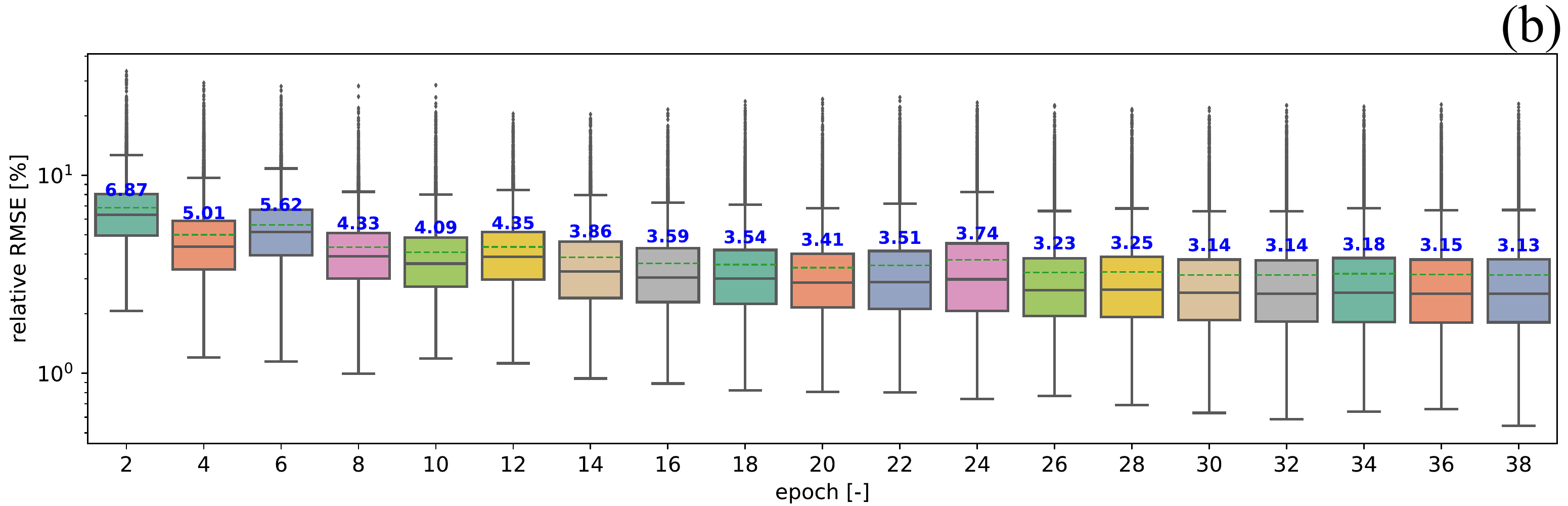}
         \includegraphics[keepaspectratio, height=4.0cm]{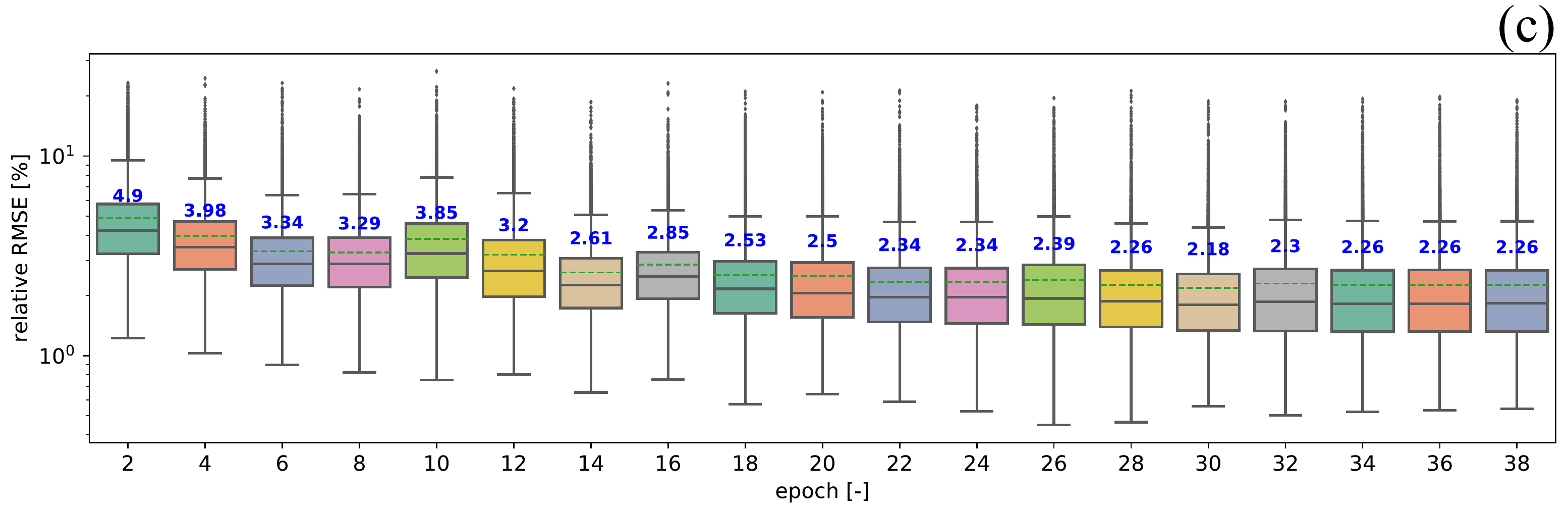}
         \includegraphics[keepaspectratio, height=4.0cm]{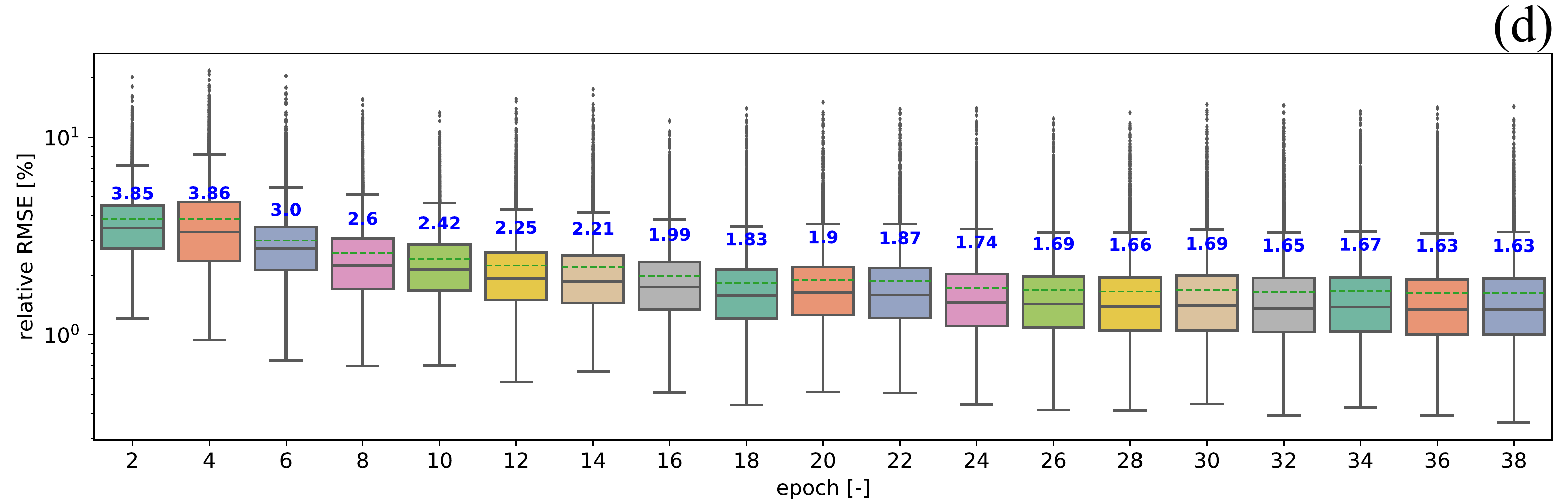}
   \caption{Example 1: relative Root Mean Square Error (relative RMSE) of ILI for (a) $\mathrm{M} = 1250$, (b) $\mathrm{M} = 2500$, (c) $\mathrm{M} = 5000$, and (d) $\mathrm{M} = 10000$. These results are calculated based on the validation set (500 samples). We note that the black dots represent outliers, and the box plot covers the interval from the 25th percentile to 75th percentile, highlighting the mean (50th percentile) with an black line. Blue line and blue text represent a mean value.}
   \label{fig:ex1_val_model2}
\end{figure}

The first Example test cases from the Zinn \& Harvey transformation are shown in Figures \ref{fig:pics_of_test_com}a-b, e-f, including $\bm{k}$ fields, FOM and ROM results, and DIFF fields for pressure and displacement fields, respectively. The box plots of relative RMSE values of pressure ($\hat{p}_h$) during training for different training samples are presented for NLI and ILI models with the validation set in Figures \ref{fig:ex1_val_model1} and \ref{fig:ex1_val_model2}, respectively. As expected, the relative RMSE values improve over epochs during training, but reaches a plateau around epochs $\approx$ 32-34. The model performance is improved with increasing the number of training samples ($\mathrm{M}$). Figures \ref{fig:ex1_val_model1} and \ref{fig:ex1_val_model2} show that the ILI model performs slightly better than the NLI model. The behavior of the results of $\hat{\bm{u}}_h$ is similar to that of $\hat{p}_h$ (results not shown). \par

The best-trained model is tested against the test set. The distributions of $\hat{p}_h$ and $\hat{\bm{u}}_h$ for the selected test cases are shown in Figure \ref{fig:pics_of_test_com}a-b and e-f, respectively. The DIFF values are very low (i.e., less than one percent of the average field values). The relative RMSE results of both $\hat{p}_h$ and $\hat{\bm{u}}_h$ of the test set are provided in Table \ref{tab:rmse_test}, which are very close to those of the validation set (Figures \ref{fig:ex1_val_model1} and \ref{fig:ex1_val_model2}). As in training, model performance improves with increasing $\mathrm{M}$. Besides, ILI always performs better than NLI. The relative RMSE of displacement ($\hat{\bm{u}}_h$) is generally lower than that of pressure ($\hat{p}_h$), which attributes to the relatively uniform response of displacement fields, compared to the pressure field as shown in Figure \ref{fig:pics_of_test_com}. Hence, the ROM can learn the solution easier.  \par

\begin{table}[!ht]
  \centering
  \caption{The relative RMSE (Eq. \eqref{eq:rmse}) results for testing data of three example cases as a function of the number of training data ($\mathrm{M}$) for pressure and magnitude of displacement (in parenthesis). Each example is evaluated with both NLI and ILI models. }
    \begin{tabular}{|c|l|c|c|c|c|}
    \hline
    \multirow{9}[0]{*}{ \makecell{\textbf{Pressure} \\  (\textbf{Displacement})}}  &\textbf{Example 1} & $\mathrm{M} = 1250$ & $\mathrm{M} = 2500$ & $\mathrm{M} = 5000$ & $\mathrm{M} = 10000$ \\
\cline{2-6}          & NLI (\%) & \cellcolor[rgb]{ .973,  .412,  .42}4.63 (4.32) & \cellcolor[rgb]{ .98,  .686,  .694}3.24 (2.98) & \cellcolor[rgb]{ .988,  .859,  .871}2.34 (2.14) & \cellcolor[rgb]{ .988,  .976,  .988}1.74 (1.57) \\
\cline{2-6}          & ILI (\%) & \cellcolor[rgb]{ .976,  .427,  .435}4.55 (4.13) & \cellcolor[rgb]{ .98,  .702,  .71}3.15 (2.78) & \cellcolor[rgb]{ .988,  .867,  .878}2.30 (2.03) & \cellcolor[rgb]{ .988,  .988,  1}1.67 (1.33) \\
\cline{2-6}          & \textbf{Example 2} & $\mathrm{M} = 1250$ & $\mathrm{M} = 2500$ & $\mathrm{M} = 5000$ & $\mathrm{M} = 10000$ \\
\cline{2-6}          & NLI (\%) & \cellcolor[rgb]{ .976,  .443,  .451}3.60 (3.60) & \cellcolor[rgb]{ .98,  .671,  .682}2.61 (2.51) & \cellcolor[rgb]{ .988,  .851,  .863}1.83 (1.47) & \cellcolor[rgb]{ .988,  .984,  .996}1.24 (1.10) \\
\cline{2-6}          & ILI (\%) & \cellcolor[rgb]{ .973,  .412,  .42}3.73 (3.37) & \cellcolor[rgb]{ .98,  .686,  .694}2.55 (2.07) & \cellcolor[rgb]{ .988,  .886,  .898}1.67 (1.26) & \cellcolor[rgb]{ .988,  .988,  1}1.22 (0.83) \\
\cline{2-6}          & \textbf{Example 3} & $\mathrm{M} = 2500$ & $\mathrm{M} = 5000$ & $\mathrm{M} = 10000$ & $\mathrm{M} = 20000$ \\
\cline{2-6}          & NLI (\%) & \cellcolor[rgb]{ .973,  .412,  .42}3.36 (3.28) & \cellcolor[rgb]{ .984,  .706,  .718}2.31 (2.28) & \cellcolor[rgb]{ .988,  .89,  .902}1.65 (1.27) & \cellcolor[rgb]{ .988,  .98,  .992}1.32 (1.27) \\
\cline{2-6}          & ILI (\%) & \cellcolor[rgb]{ .976,  .494,  .502}3.07 (2.74) & \cellcolor[rgb]{ .984,  .725,  .737}2.24 (2.15) & \cellcolor[rgb]{ .988,  .894,  .906}1.63 (1.18) & \cellcolor[rgb]{ .988,  .988,  1}1.29 (1.05) \\
    \hline
    \end{tabular}%
    \footnotesize{   Example 3: a total number of $\mathrm{M}$ is the sum of training data from both Examples 1 and 2.}
  \label{tab:rmse_test}%
\end{table}%

\subsection{Example 2: bimodal transformation}

\begin{figure}[!ht]
   \centering
         \includegraphics[keepaspectratio, height=4.0cm]{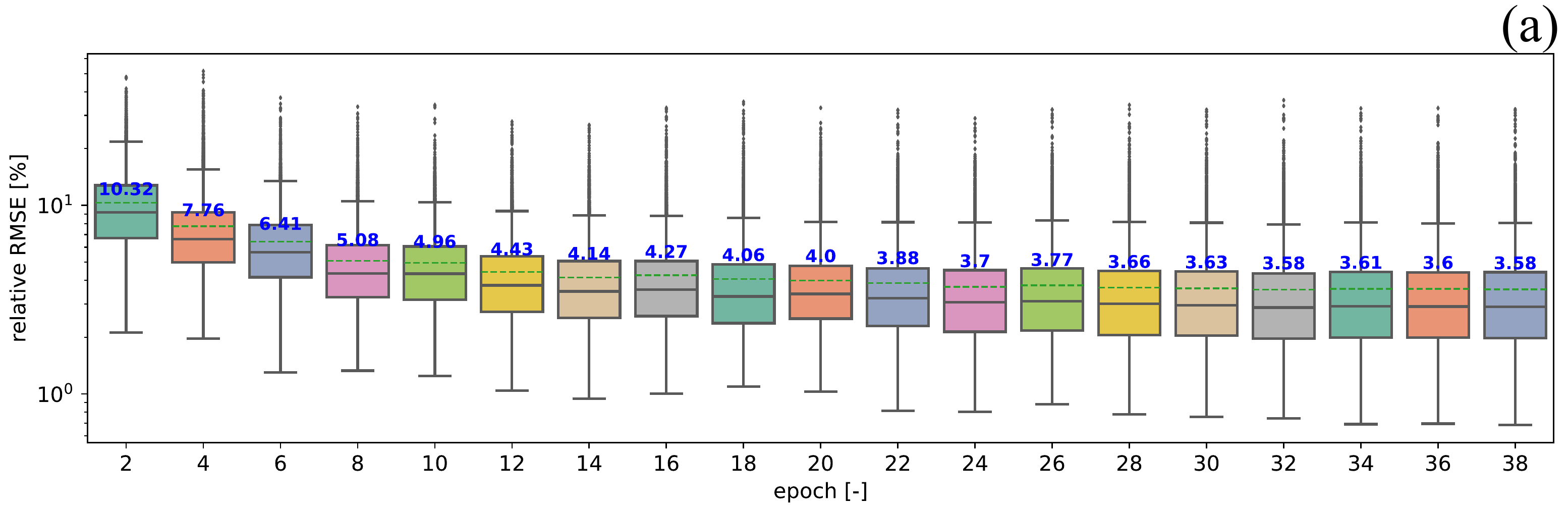}
         \includegraphics[keepaspectratio, height=4.0cm]{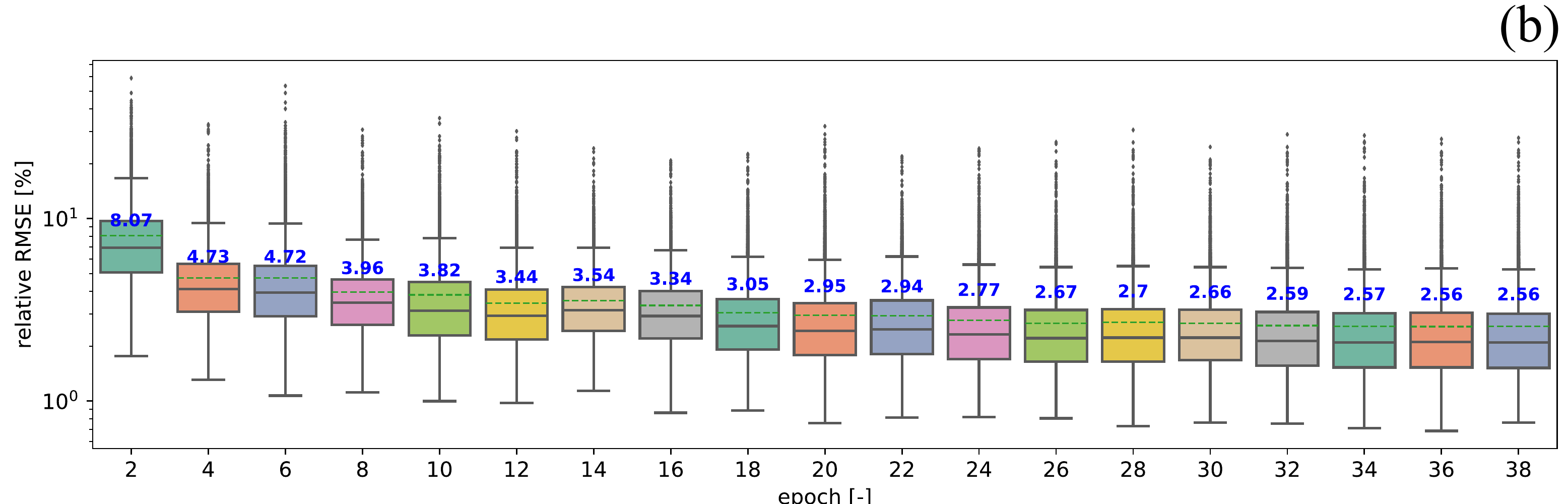}
         \includegraphics[keepaspectratio, height=4.0cm]{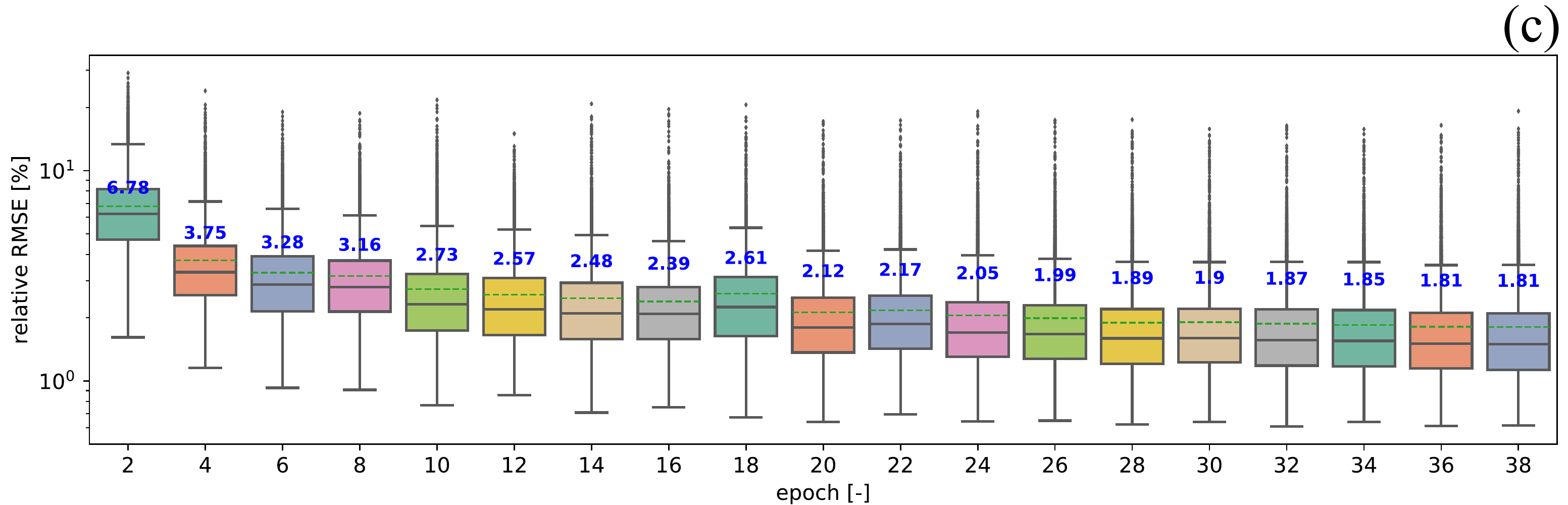}
         \includegraphics[keepaspectratio, height=4.0cm]{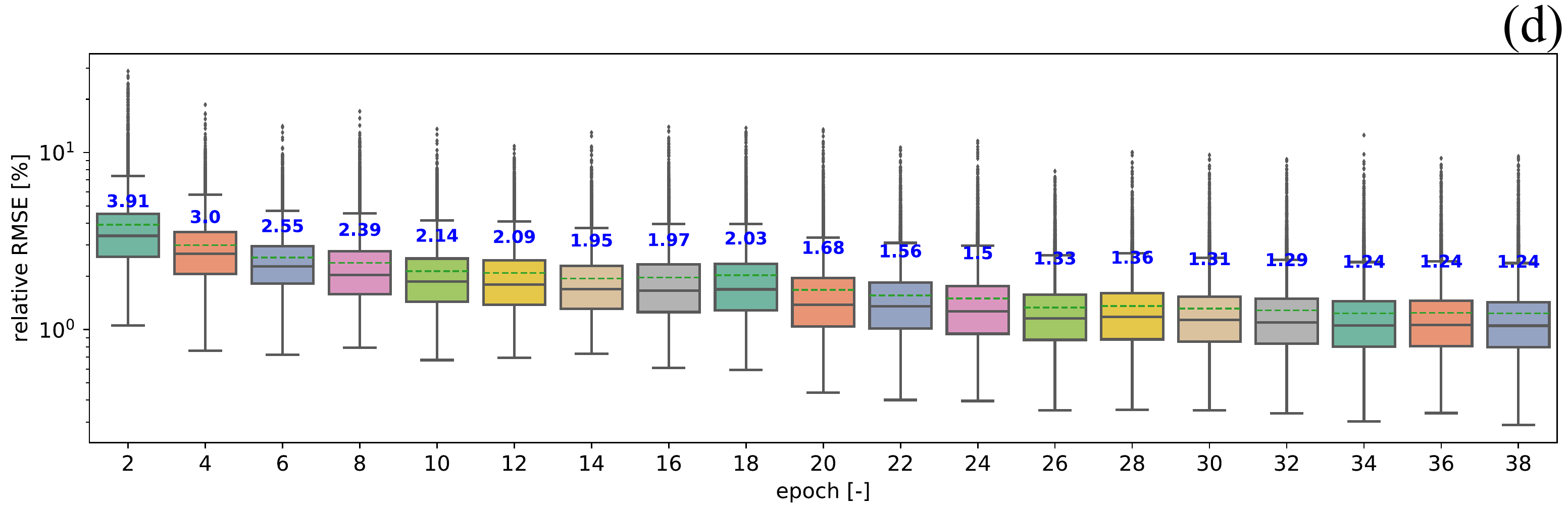}
   \caption{Example 2: relative Root Mean Square Error (relative RMSE) of NLI for (a) $\mathrm{M} = 1250$, (b) $\mathrm{M} = 2500$, (c) $\mathrm{M} = 5000$, and (d) $\mathrm{M} = 10000$. These results are calculated based on the validation set (500 samples). We note that the black dots represent outliers, and the box plot covers the interval from the 25th percentile to 75th percentile, highlighting the mean (50th percentile) with an black line. Blue line and blue text represent a mean value.}
   \label{fig:ex2_val_model1}
\end{figure}

\begin{figure}[!ht]
   \centering
         \includegraphics[keepaspectratio, height=4.0cm]{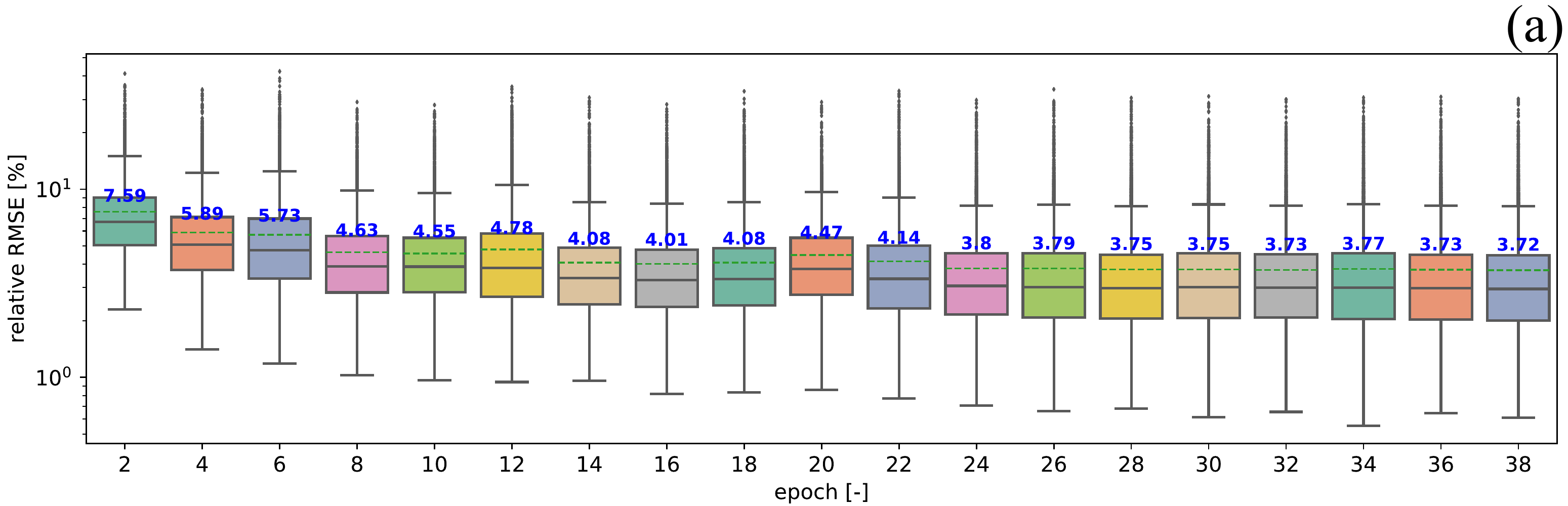}
         \includegraphics[keepaspectratio, height=4.0cm]{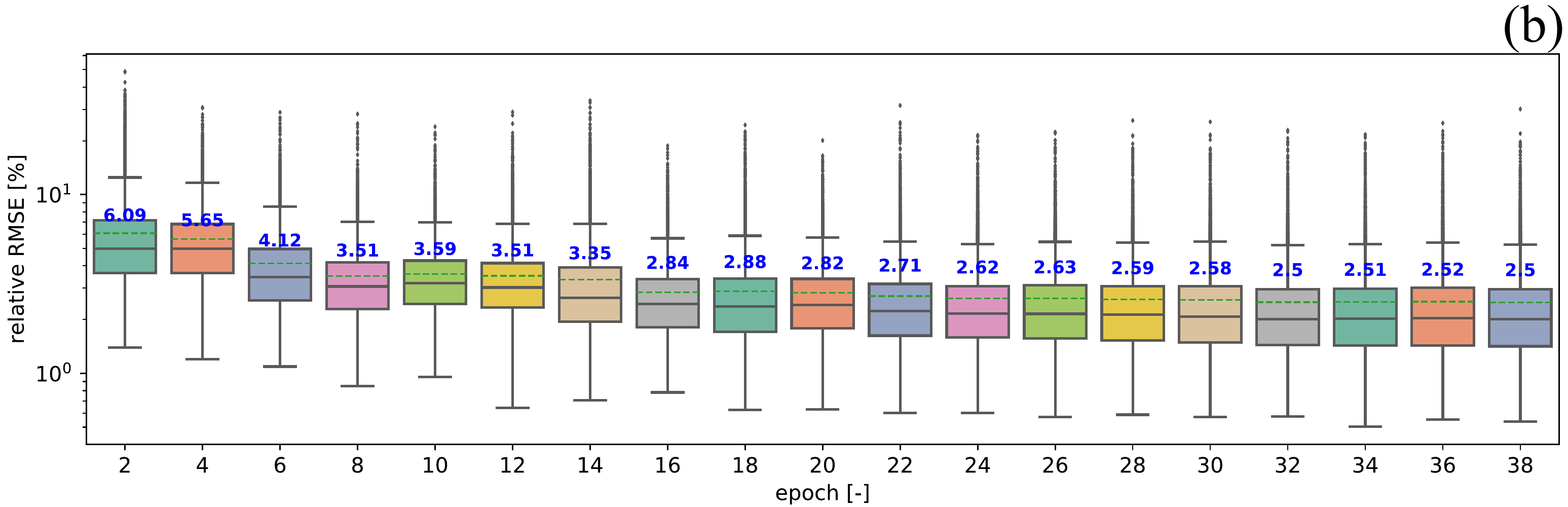}
         \includegraphics[keepaspectratio, height=4.0cm]{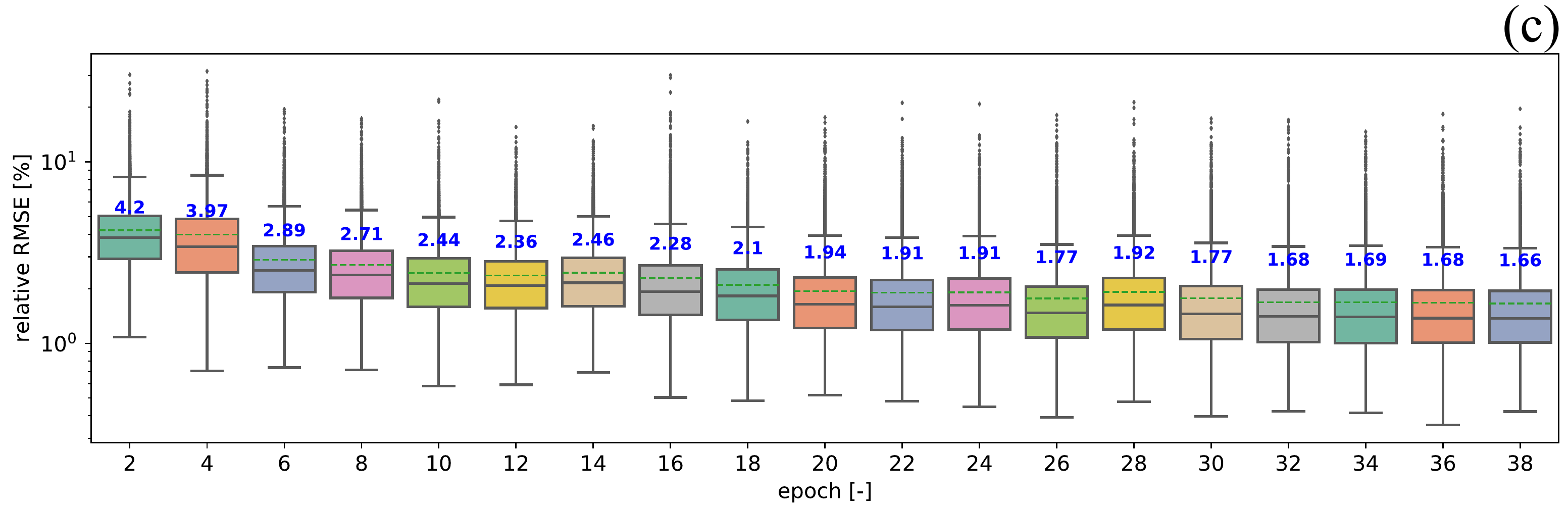}
         \includegraphics[keepaspectratio, height=4.0cm]{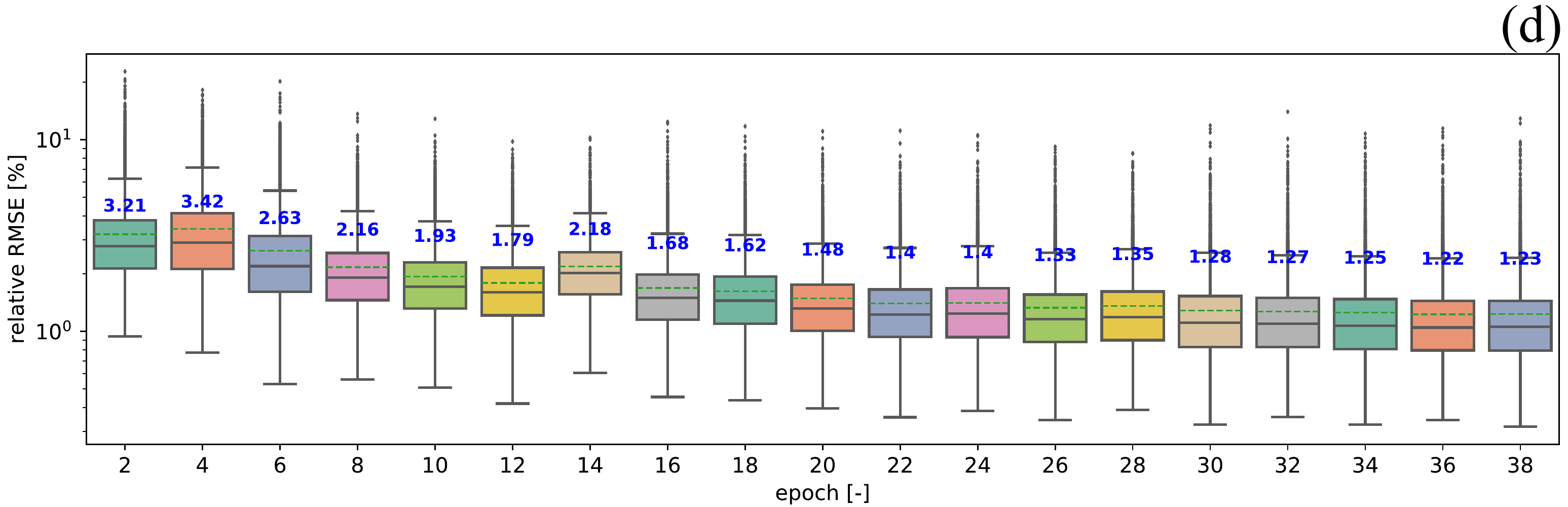}
   \caption{Example 2: relative Root Mean Square Error (relative RMSE) of ILI for (a) $\mathrm{M} = 1250$, (b) $\mathrm{M} = 2500$, (c) $\mathrm{M} = 5000$, and (d) $\mathrm{M} = 10000$. These results are calculated based on the validation set (500 samples). We note that the black dots represent outliers, and the box plot covers the interval from the 25th percentile to 75th percentile, highlighting the mean (50th percentile) with an black line. Blue line and blue text represent a mean value.}
   \label{fig:ex2_val_model2}
\end{figure}

The second Example presents the model performance using $\bm{k}$ fields from the bimodal transformation, which has a narrow range of $\bm{k}$ values with wider high permeability pathways. As in Example 1, selected test cases of $\bm{k}$ fields, FOM and ROM results, and DIFF fields are presented in Figure \ref{fig:pics_of_test_com}c-d and g-h. The box plot of relative RMSE values of the validation set is presented in Figures \ref{fig:ex2_val_model1} and \ref{fig:ex2_val_model2}. Similar to Example 1, the model performance improves with increasing $\mathrm{M}$. Moreover, the higher the number of epochs, the model tends to provide more accurate results. Although there are some fluctuations of the relative RMSE values at a later training stage, the model accuracy tends to improve as the training progresses. Except for the case where $\mathrm{M} = 1250$, the ILI model provides better results than the NLI. \par 

For the testing set, the ILI model provides better accuracy than the NLI except $\mathrm{M} = 1250$ for $\hat{p}_h$ (Table \ref{tab:rmse_test}). It is noted that ILI always performs better than NLI for the state variable $\hat{\bm{u}}_h$. Moreover, the relative RMSE results of $\hat{\bm{u}}_h$ is always lower than those of $\hat{p}_h$, which are similar to Example 1. \par

The relative RMSE of Example 2 is slightly lower than that of Example 1 ( Table \ref{tab:rmse_test}), however, the trend of the relative RMSE values between NLI and ILI is similar in both Examples 1 and 2. We will discuss the performance of NLI and ILI in the next sections. Since $\bm{k}$ fields from the bimodal transformation have a narrower range and wider permeable pathways with less contrast compared to those from the Zinn \& Harvey transformation (see Figure \ref{fig:pics_of_test_com}), the corresponding pressure field may have similar features. This can be seen in the DIFF distribution where the DIFF values are larger along high-pressure gradient regions in all pressure cases (Figure \ref{fig:pics_of_test_com}a-d). At $t=25$ in Example 2 (Figure \ref{fig:pics_of_test_com}c), high DIFF values are mostly located near the top boundary where the pressure boundary is set to zero after the initial pressure of 1000 Pa. Over time the DIFF distribution propagates as the pressure contrast migrates along the high permeability regions (e.g., the DIFF fields at $t=250$ in Figure \ref{fig:pics_of_test_com}d). Compared to Example 1 where pressure gradients tend to be slightly gradual (e.g., wider transition) and the high DIFF values are distributed in larger areas, Example 2 cases show the higher contrast of pressure values along with the pressure transition, and the high DIFF values tend to be distributed more locally (Figure \ref{fig:pics_of_test_com}a-d). \par

Although this comparison qualitatively shows the dependency of the model performance on input and output characteristics, it is also well-known that deep neural networks often need complex neural network architecture to extract and learn high-frequency features such as high permeability contrast and high-pressure gradients in this work \citep{xu2019frequency}. A recent work by \cite{kim2021connectivity} transformed physical connectivity information of the high contrast drainage network into multiple binary matrices, which improved the network generation using deep convolutional GAN. However, the success rates of the drainage network with proper connectivity were relatively low. CcGAN developed in this work shows that although it is still challenging to improve the prediction accuracy, the increase in the training data sets may provide a potential solution to this challenging problem. However, the increase of training data sets also increases required computational resources. This aspect will be discussed later. For displacement results (Figure \ref{fig:pics_of_test_com}e-h) the relative RMSE results in Example 2 (Table \ref{tab:rmse_test}) follow the same trend in the pressure results. The lower relative RMSE values of displacement than pressure also stem from the smooth displacement fields compared to pressure fields. It is noted that the relative RMSE values of the test set are similar to those of the validation set. \par

\subsection{Example 3: Combined Zinn \& Harvey and bimodal transformations}

\begin{figure}[!ht]
   \centering
         \includegraphics[keepaspectratio, height=4.0cm]{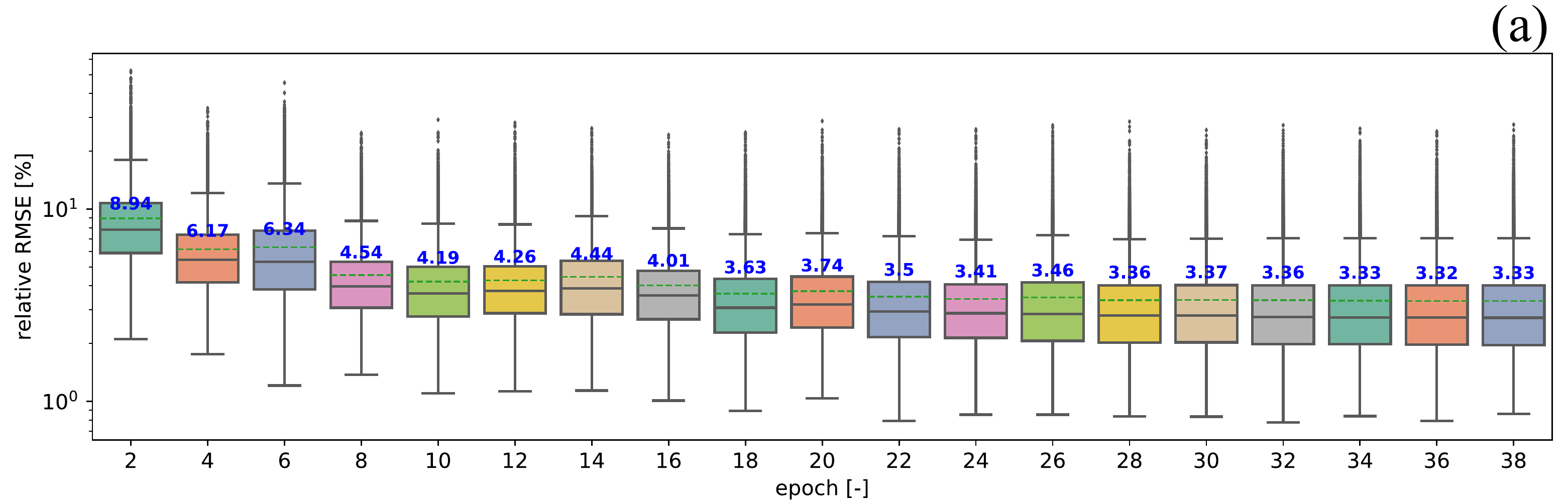}
         \includegraphics[keepaspectratio, height=4.0cm]{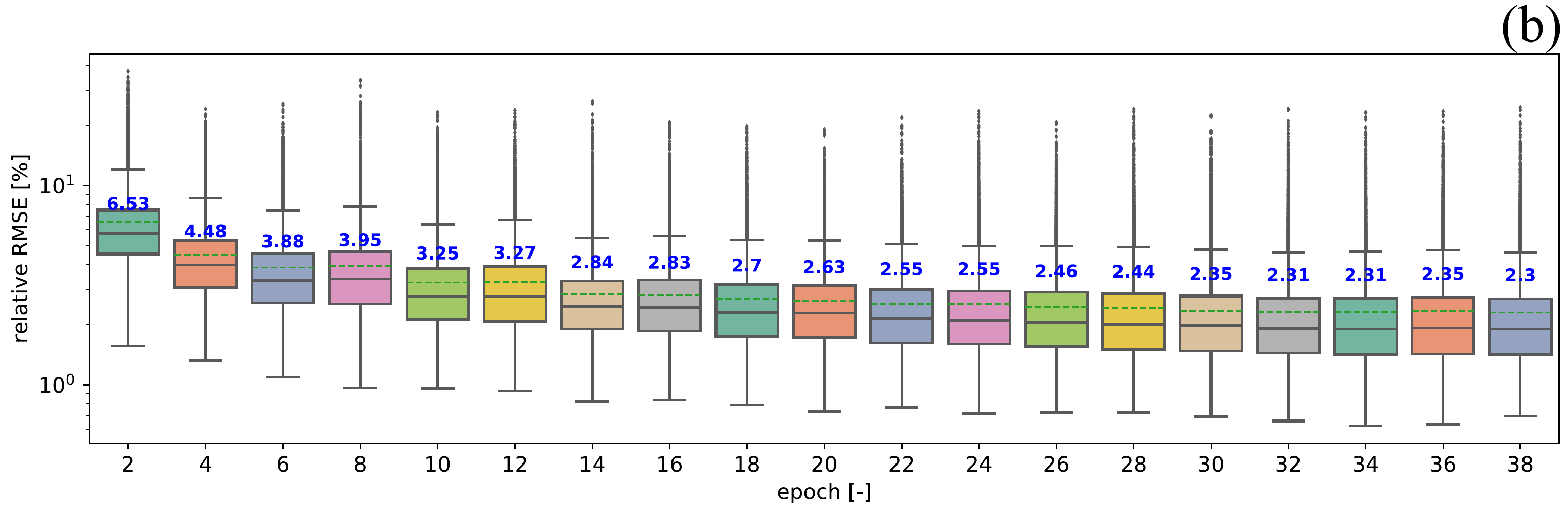}
         \includegraphics[keepaspectratio, height=4.0cm]{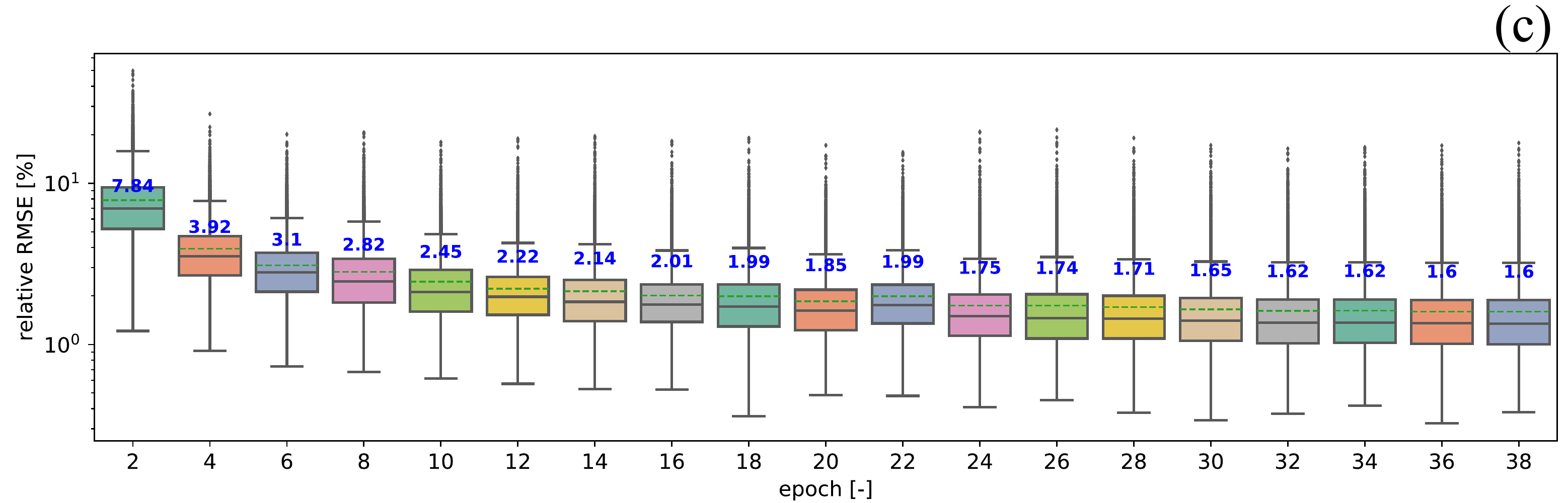}
         \includegraphics[keepaspectratio, height=4.0cm]{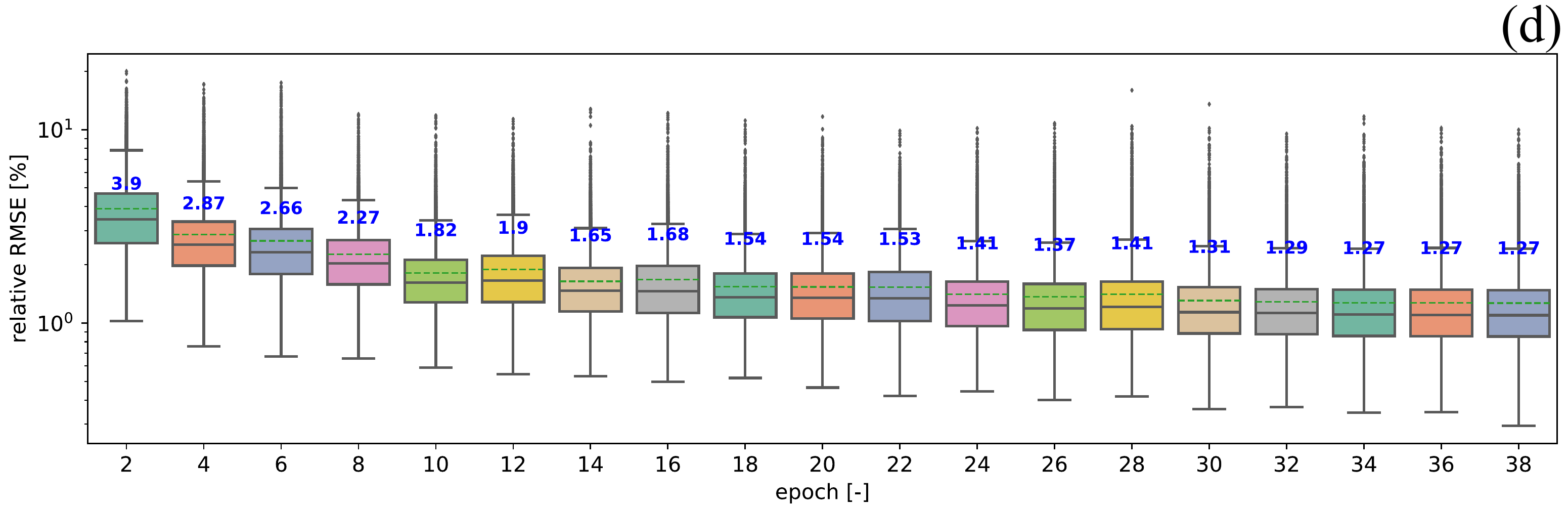}
   \caption{Example 3: relative Root Mean Square Error (relative RMSE) of NLI for (a) $\mathrm{M} = 2500$, (b) $\mathrm{M} = 5000$, (c) $\mathrm{M} = 10000$, and (d) $\mathrm{M} = 20000$. These results are calculated based on the validation set (500 samples). We note that the black dots represent outliers, and the box plot covers the interval from the 25th percentile to 75th percentile, highlighting the mean (50th percentile) with an black line. Blue line and blue text represent a mean value.}
   \label{fig:ex3_val_model1}
\end{figure}

\begin{figure}[!ht]
   \centering
         \includegraphics[keepaspectratio, height=4.0cm]{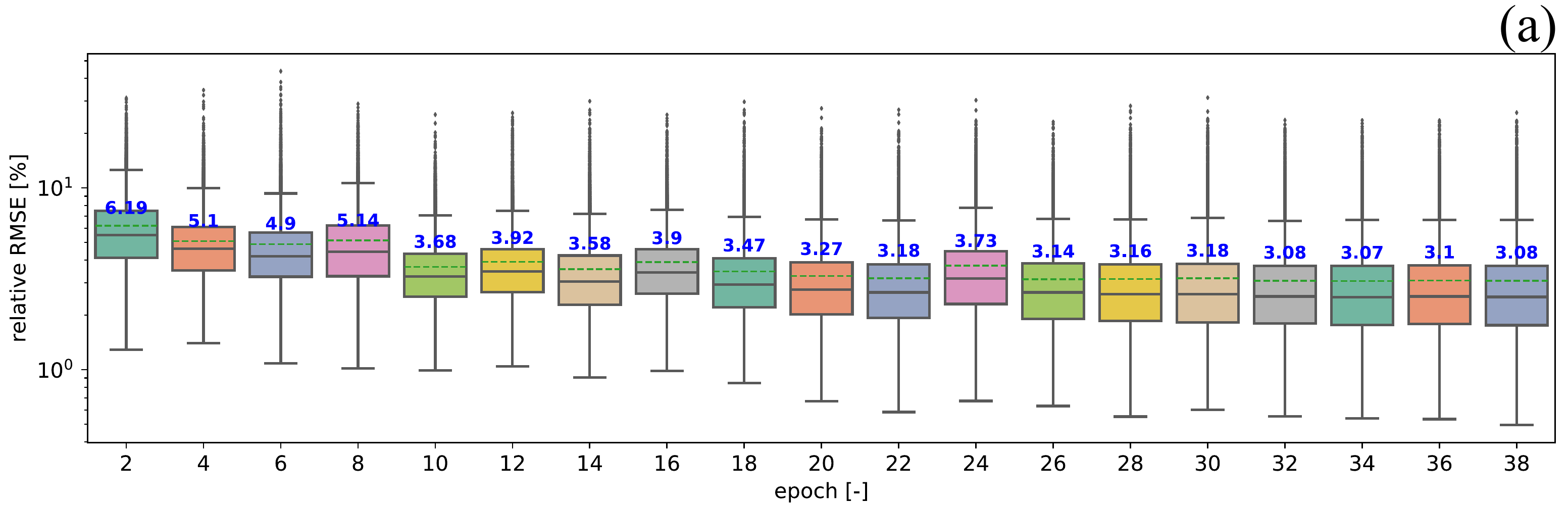}
         \includegraphics[keepaspectratio, height=4.0cm]{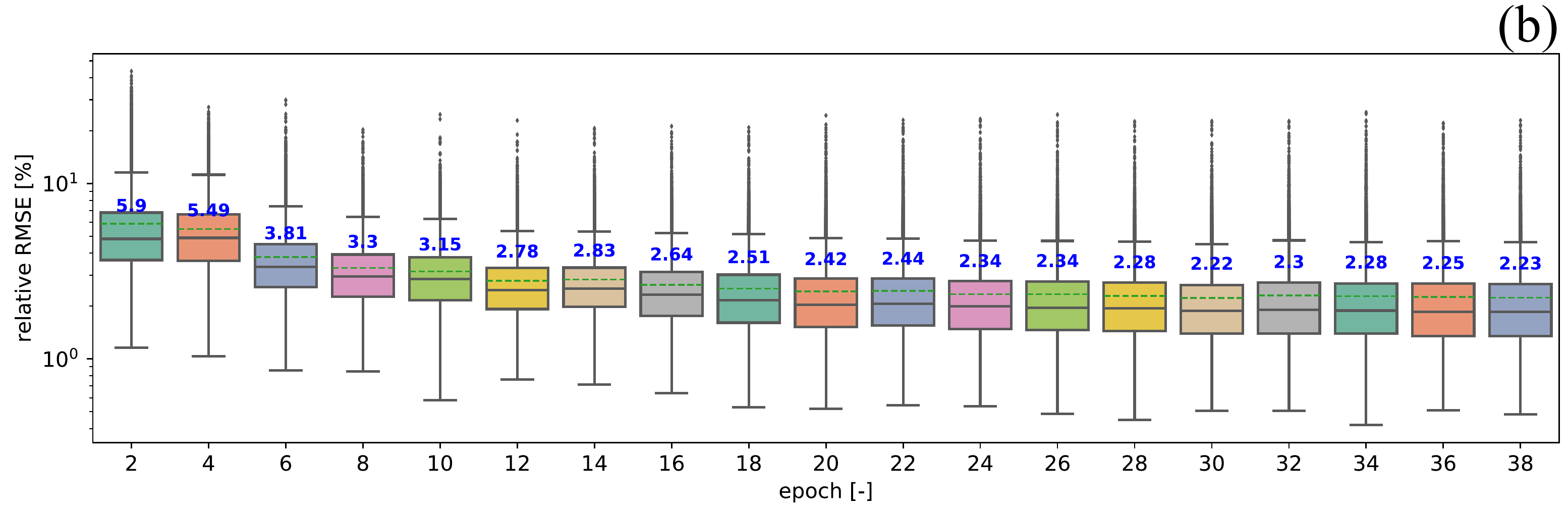}
         \includegraphics[keepaspectratio, height=4.0cm]{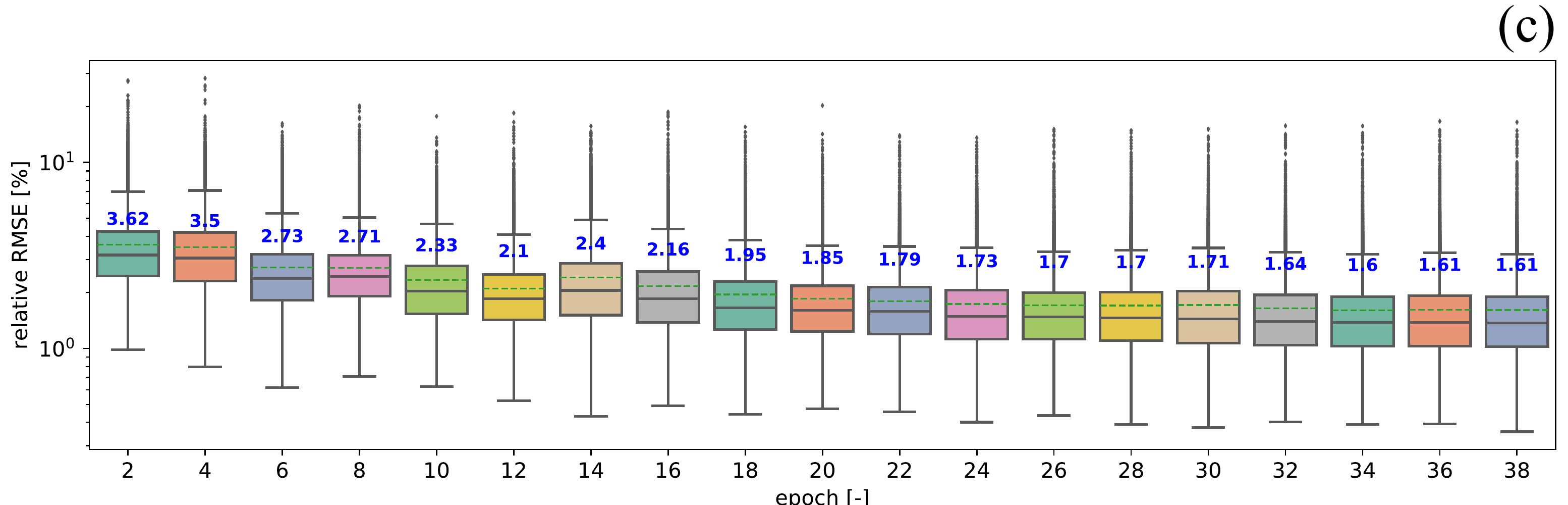}
         \includegraphics[keepaspectratio, height=4.0cm]{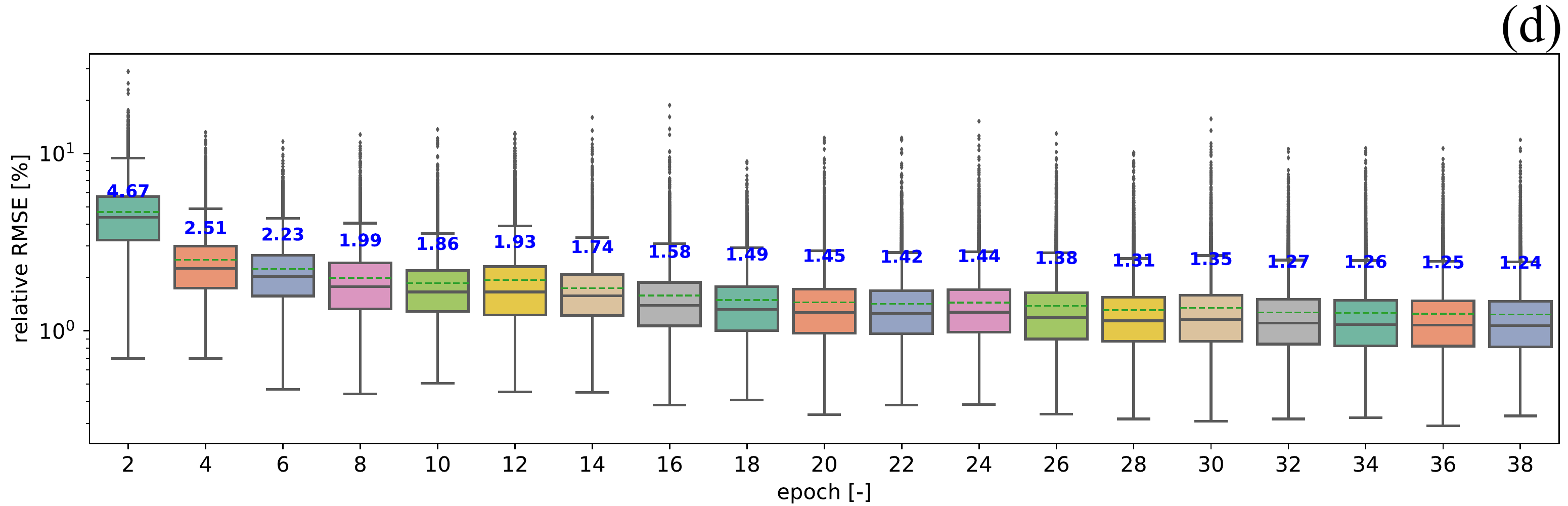}
   \caption{Example 3: relative Root Mean Square Error (relative RMSE) of ILI for (a) $\mathrm{M} = 2500$, (b) $\mathrm{M} = 5000$, (c) $\mathrm{M} = 10000$, and (d) $\mathrm{M} = 20000$. These results are calculated based on the validation set (500 samples). We note that the black dots represent outliers, and the box plot covers the interval from the 25th percentile to 75th percentile, highlighting the mean (50th percentile) with an black line. Blue line and blue text represent a mean value.}
   \label{fig:ex3_val_model2}
\end{figure}

In Example 3, permeability fields from both Zinn \& Harvey and bimodal transformations are used to test the generalization ability of the proposed approach (i.e., Figures \ref{fig:pics_of_test_com}a-h). As discussed earlier, Example 3 can represent different types of heterogeneity with high permeable pathways. The relative RMSE of the validation set for pressure fields ($\hat{p}_h$) is presented in Figures \ref{fig:ex3_val_model1} and \ref{fig:ex3_val_model2}. Similar to Examples 1 and 2, the model accuracy improves with increasing $\mathrm{M}$, and we did not observe any over-fitting behavior. \par

As presented in Table \ref{tab:rmse_test}, for both pressure and displacement fields, the ILI model performs better than the NLI, and the model with a higher $\mathrm{M}$ (i.e., more training data) provides better accuracy. Note that Example 3 has two times higher $\mathrm{M}$ than the other two cases. Overall, the relative RMSE values in Example 3 are closer to those in Example 1 rather than Example 2 for the same number of $\mathrm{M}$. This indicates that more challenging fields for ML training predominantly govern the model performance with combined fields. In addition, Example 3 tends to have higher relative RMSE values than Examples 1 and 2 for the lower number of $\mathrm{M}$ (e.g., $\mathrm{M}$ = 2500 and 5000 for pressure and displacement, respectively). As the number of $\mathrm{M}$ increases, however, Example 3 has lower RMSE values than Example 1 and gets closer to Example 2, indicating that more training data sets improve the ML model to a certain degree. Although there may be more optimal hyperparameters to train both fields better, these results demonstrate the general learning capability of the proposed models. The inclusion of a more challenging data set will increase the generalization capability of the trained model. \par

\subsection{Computational costs} 
A summary of the wall time used for each operation (i.e., steps 2 to 4 in Figure \ref{fig:intro}) is presented in Table S4\ref{tab:wall_time} where the time for step 1 or initialization was negligible compared to the other steps. The FE model (FOM) was run using a single AMD Ryzen Threadripper 
3970X, while training and testing of CcGAN or ROM models were carried out with single GPU (NVIDIA Quadro RTX 6000). With the fixed number of $N^t = 10$ throughout this study, each FOM simulation (per $\bm{\mu}^{(i)}$) takes about 40 seconds. Consequently, if we select $\mathrm{M} = 10000$ as an example, it would take about 400,000 seconds ($\approx$111 hrs) to build the training set. To train the model using $\mathrm{M} = 10000$, it takes approximately 30 hours. During the online or the prediction phase, however, the trained model could provide its result within 0.001 seconds per testing $\left(t^n, \bm{\mu}^{(i)} \right)$. It should be noted that the ROM as the trained model in this work is not required to have the number of $N^t$ in which FOM uses. Assuming the ROM also uses $N^t = 10$, it still provides a speed-up by 10,000 times. One advantage of the ROM framework compared to the FOM is that it can also deliver the solution at any time, including times that do not exist in the training snapshots since it is simply a nonlinear interpolation in output space. This characteristic is an asset of the ROM in this work because it is not bound by any time-stepping constraints and can evaluate quantities of interest at any time required. For instance, we may be interested in pressure and displacement fields at one, two, and three hours with given different $\bm{\mu}$. To achieve this with FOM, it may need to go through many steps in between. However, ROM enables us to evaluate it at those three times only.

\begin{table}[!ht]
\centering
\caption{Comparison of the wall time (seconds) used for each operation presented in Figure 1 (main text). $\bm{\mu}$ is a set of parameterize spatial fields, and $\bm{\mu}_i \in \bm{\mu}$.}
\begin{tabular}{|l|c|c|c|}
\hline
                             & NLI & ILI & remark   \\ \hline
Build FOM snapshots                          & 40    & 40     & per $\bm{\mu}_i$ for $N^t = 10$  \\ \hline
Train ROM with $\mathrm{M} = 1250$                          & 12600    & 12600    & approximately 3.75 hours    \\ \hline
Train ROM with $\mathrm{M} = 2500$                          & 25200    & 25200    & approximately 7.5 hours    \\ \hline
Train ROM with $\mathrm{M} = 5000$                          & 50400    & 50400    & approximately 15 hours    \\ \hline
Train ROM with $\mathrm{M} = 10000$                          & 108000    & 108000    & approximately 30 hours    \\ \hline
Train ROM with $\mathrm{M} = 20000$                          & 216000    & 216000    & approximately 60 hours    \\ \hline
Prediction   & 0.001     & 0.001       & per testing $\left(t^n, \bm{\mu}_i \right) $ \\ \hline
\end{tabular}
\label{tab:wall_time}
\end{table}

\subsection{Prediction accuracy and subsurface physics applications}

With three examples presented, we observe that the ILI model performs better than the NLI for all cases except one particular case (Table \ref{tab:rmse_test}). This finding is in good agreement with classification problems in \cite{ding2020continuous}. \cite{ding2020continuous} speculates that the ILI overcomes the label inconsistency of the classification problems while the NLI could not. Our reasoning for the outperformance of the INI stems from continuous conditional batch normalization, which can carry out temporal information more consistently than the element-wise addition in the NLI model. In addition to the skip connections that are essential in transferring multiscale features between input (permeability fields) and output (pressure and displacement fields) (see \cite{kadeethum2021framework}), CcGAN provides the framework to account for temporal features over time, resulting in better representation of temporal patterns. \par
 
One key aspect of many subsurface energy applications is a coupled process that involves hydrogeology and geomechanics. Although ML-based data-driven modeling has been increasingly studied for reservoir modeling, most are still limited to uncoupled processes (e.g., \cite{lange2020machine,miah2020predictive}) or relatively homogeneous fields (e.g., \cite{zounemat2021ensemble,zhao2021reduced}). The CcGAN approach proposed in this study demonstrates its capability to handle coupled hydro-mechanical processes with relative RMSE less than 2\% of the transient pressure and displacement responses in the worst case. The results also show that the relative RMSE of the ILI model can be improved to a scale of about 1\% with more training data sets, which can be acceptable given the uncertainty and observational error in the subsurface systems. Additionally, our framework for model prediction (i.e., online stage) achieves up to $\approx$10,000x speed-up compared to the FE solver. Note that the computational advantage of ML-driven ROM models will increase further with increasing degree of freedom in the FE solver (e.g., three-dimensional and longer transient problems). This computational advantage and accuracy will enable us to achieve real-time reservoir management and robust uncertainty quantification even for vast parameter space. At the same time, the ROM can be updated offline as necessary. It should be noted that the method presented here can be extended to incorporate more continuous variables into the system. For instance, besides the time domain, we could also add Young's modulus and Poisson ratio into the CcGAN model. Furthermore, since this model is \emph{data-driven}, it is not limited to only coupled hydro-mechanical processes presented in this manuscript but also applicable to other coupled processes such as thermal-hydrological-mechanical-chemical (THMC) processes. \par

\section{Conclusions} \label{sec:conclusions}

This work presents a data-driven framework for solving a system of time-dependent partial differential equations, more explicitly focusing on coupled hydro-mechanical processes in heterogeneous porous media. This framework can be used as a proxy for time-dependent coupled processes in heterogeneous porous media, which is challenging in classical model order reduction. Our framework is developed upon continuous conditional generative adversarial networks (CcGAN) composed of the U-Net generator and patch-based critic. The model has two variations: (1) the time domain is introduced to only the generator's bottleneck using element-wise addition (i.e., NLI), and (2) the time domain is injected into all layers inside the generator through conditional batch normalization (i.e., ILI). The critic is similar for both models. Our approach is desirable because it does not require any cumbersome modifications of FOM source codes and can be applied to any existing FOM platforms. In this regard, the CcGAN approach to solve time-dependent PDEs is uniquely employed to develop a data-driven surrogate model given highly heterogeneous permeability fields. We illustrate that our framework could efficiently and accurately approximate finite element results given a wide variety of permeability fields. Our results have a relative root mean square error of less than 2\% with 10,000 samples for training. Additionally, it could speed up at least 10,000 times compared to a forward finite element solver. ILI delivers slightly better results than NLI without any observable additional computational costs. To this end, this framework will enable us to do a large-scale operation of real-time reservoir management or uncertainty quantification with complex heterogeneous permeability fields, which are practically very difficult to do with FOM and traditional model order reductions.

\section{Acknowledgments}
TK and HY were supported by the Laboratory Directed Research and Development program at Sandia National Laboratories and US Department of Energy Office of Fossil Energy and Carbon Management, Science-Informed Machine Learning to Accelerate Real Time Decisions-Carbon Storage (SMART-CS) initiative. 
DO acknowledges support from Los Alamos National Laboratory's Laboratory Directed Research and Development Early Career Award (20200575ECR). 
YC acknowledges LDRD funds (21-FS-042) from Lawrence Livermore National Laboratory. Lawrence Livermore National Laboratory is operated by Lawrence Livermore National Security, LLC, for the U.S. Department of Energy, National Nuclear Security Administration under Contract DE-AC52-07NA27344 (LLNL-JRNL-827590).
HV is grateful to the funding support from the U.S. Department of Energy (DOE) Basic Energy Sciences (LANLE3W1).
NB acknowledges startup support from the Sibley School of Mechanical and Aerospace Engineering, Cornell University.
Sandia National Laboratories is a multimission laboratory managed and operated by National Technology and Engineering Solutions of Sandia, LLC., a wholly owned subsidiary of Honeywell International, Inc., for the U.S. Department of Energy’s National Nuclear Security Administration under contract DE-NA-0003525. This paper describes objective technical results and analysis. Any subjective views or opinions that might be expressed in the paper do not necessarily represent the views of the U.S. Department of Energy or the United States Government.

\section{Data and code availability}
We will release our CcGAN scripts and training and testing data to reproduce all results in the manuscript through the Sandia National Laboratories software portal — a hub for GitHub-hosted open source projects \hfill \break (\url{https://github.com/sandialabs}) after the manuscript will be accepted.  
For review purpose we provide the training, validation, and testing data or the scripts used to generate them at \url{https://codeocean.com/capsule/2052868/tree/v1} \cite{kadeethum2021frameworksourcecode}. FOM source codes were already made available here: \url{https://github.com/teeratornk/jcp_YJCPH_110030_git} as well as a tutorial (see tutorial number 9) for \textbf{multiphenics} package \url{https://github.com/multiphenics}.

\newpage

\begin{appendices}

\beginapp

\section{Problem description, model geometry, and boundaries} \label{sec:prob_description}

To recap, the coupled HM processes in porous media are governed by two equations. The first equation is linear momentum balance equation

\begin{equation} \label{eq:linear_balance}
\begin{split}
\nabla \cdot \bm{\sigma}^{\prime}(\bm{u}) -\alpha \nabla \cdot \left(p \mathbf{I}\right)
+ \bm{f} = \bm{0}  &\text { \: in \: } \Omega \times \mathbb{T}, \\
\bm{u} =\bm{u}_{D} &\text { \: on \: } \partial \Omega_{D} \times \mathbb{T},\\
\bm{\sigma} {(\bm{u})} \cdot  \mathbf{n}=\bm{t_D} &\text { \: on \: } \partial \Omega_{N} \times \mathbb{T}, \\
\bm{u}=\bm{u}_{0}  &\text { \: in \: } \Omega \text { at } t = 0,
\end{split}
\end{equation}

\noindent
where $\bm{\sigma}^{\prime}$ is the effective Cauchy stress tensor, $p$ is the pore pressure, $\bm{u}$ is bulk displacement, $\bm{u}_0$ is an initial displacement, $\mathbf{I}$ is the second-order identity tensor, $\alpha$ is the Biot coefficient, $\bm{f}$ is the body force term defined as $\rho \phi \mathbf{g}+\rho_{s}(1-\phi) \mathbf{g}$, where $\rho$ is the fluid density, $\rho_s$ is the solid density, $\phi$ is the porosity, $\mathbf{g}$ is the gravitational acceleration vector, $\bm{u}_D$ and ${\bm{t}_D}$ are prescribed displacement and traction values at the boundaries, respectively, and $\mathbf{n}$ is the unit normal vector to the boundary. \par 

The second equation is mass balance equation read

\begin{equation} \label{eq:mass_balance}
\begin{split}
\left(\frac{1}{M}+\frac{\alpha^{2}}{K}\right) \frac{\partial p}{\partial t}+\frac{\alpha}{K} \frac{\partial \sigma_{v}}{\partial t}- \nabla \cdot \left( \bm{\kappa} \nabla p \right)= g  &\text { \: in \: } \Omega \times \mathbb{T}, \\
p=p_{D} &\text { \: on \: } \partial \Omega_{D} \times \mathbb{T}, \\
- \bm{\kappa} \nabla p \cdot \mathbf{n}=q_{D} &\text { \: on \:} \partial \Omega_{N} \times \mathbb{T}, \\
p=p_{0} &\text { \: in \: } \Omega \text { at } t = 0,
\end{split}
\end{equation}

\noindent
where is the Biot modulus, $\sigma_{v}:=\frac{1}{3} \tr(\bm{\sigma})$ is the volumetric stress, $p_D$ and $q_D$ are the given boundary pressure and flux, respectively,$p_0$ is an initial pressure, $K$ is bulk modulus, $\bm{\kappa}=\frac{\bm{k}}{\mu_f}$ is the porous media conductivity, $\mu_f$ is the fluid viscosity, $\bm{k}$ is the matrix permeability tensor defined as

\begin{equation} \label{eq:permeability_matrix}
\bm{k} :=
\begin{cases}
 \left[ \begin{array}{lll}{{k}_{xx}} & {{k}_{xy}} & {{k}_{xz}} \\ {{k}_{yx}} & {{k}_{yy}} & {{k}_{yz}} \\ {{k}_{zx}} & {{k}_{zy}} & {k}_{zz}\end{array}\right] & \text{if} \ d = 3, \ \\ \\
 \left[ \begin{array}{ll}{{k}_{xx}} & {{k}_{xy}}  \\ {{k}_{yx}} & {{k}_{yy}} \\ \end{array}\right]  & \text{if} \ d = 2, \ \\ \\
 \ {k}_{xx}  & \text{if} \ d = 1.
\end{cases}
\end{equation}

\noindent
To simplify our problem, we assume all off-diagonal terms of \eqref{eq:permeability_matrix} to be zero and all diagonal terms have similar value.

It is noted that, throughout this study, our model takes heterogeneous permeability fields, $\bm{k}$, which in general sense is $\bm{\mu}$, $\bm{\mu}_\mathrm{validation}$, or $\bm{\mu}_\mathrm{test}$ depending we are dealing with train, validation, or test sets and time, $t^n$, as input. The framework will deliver $\bm{u}_h(t^n, \bm{k}^{(i)})$ and $p_h(t^n, \bm{k}^{(i)})$, which are $\bm{X}_h(t^n, \bm{\mu}^{(i)})$ in general terms. These $\bm{u}_h(t^n, \bm{k}^{(i)})$ and $p_h(t^n, \bm{k}^{(i)})$ are approximation of $\bm{u}$ and $p$ through FOM. Please refer to these FOM source codes were already made available here: \url{https://github.com/teeratornk/jcp_YJCPH_110030_git} as well as a tutorial (see tutorial number 9) for \textbf{multiphenics} package \url{https://github.com/multiphenics}.  \par

\begin{figure}[!ht]
   \centering
    \includegraphics[width=6.0cm,keepaspectratio]{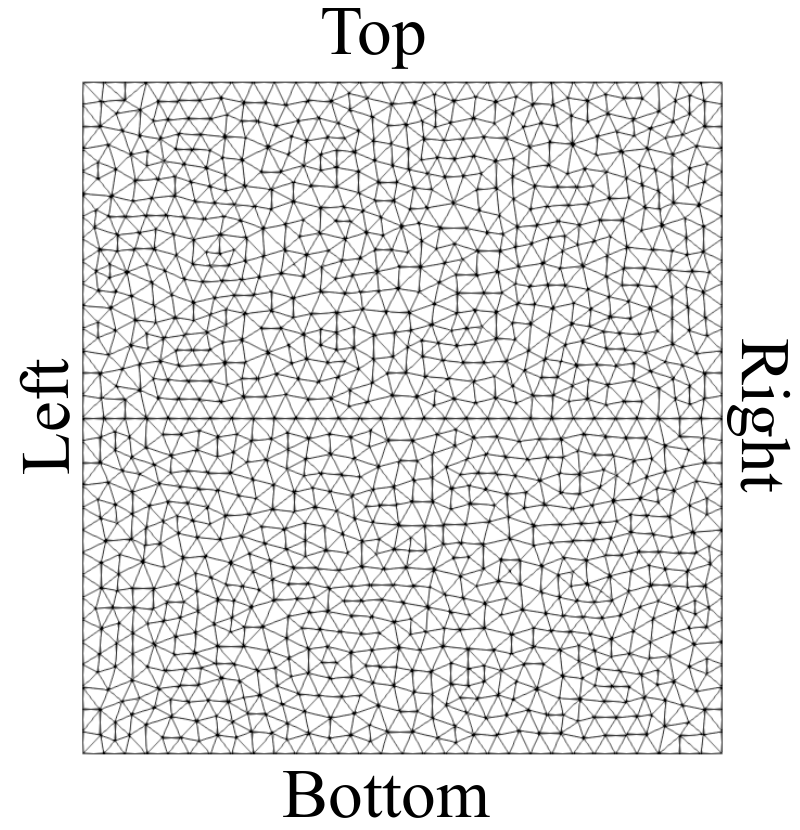} 
   \caption{Domain, its boundaries, and mesh used for all numerical examples.}
   \label{fig:mesh}
\end{figure}

The mesh and boundaries are presented in Figure \ref{fig:mesh}.

\section{Detailed information of generators and critic}

The generator resembles the well-established architecture of the U-net, which is typically used for image segmentation. The first component of the generator is a contracting block that performs two convolutions followed by a max pool operation. The second component is an expanding block in which it performs an upsampling, a convolution, and a concatenation of its two inputs. The generator used in this study consists of six contracting and six expanding blocks. Note that each contracting block uses LeakyReLU with a negative slope of 0.2 as its activation function, each expanding block uses ReLU as its activation function. The $1^{\mathrm{st}}$ convolutional layer is used to map the input channel ($\mathrm{C}_{\mathrm{in}}$) to hidden layer size ($\mathrm{H}$), and it is not subject to an activation function. The $2^{\mathrm{nd}}$ convolutional layer is used to map hidden layer size ($\mathrm{H}$) to output channel ($\mathrm{C}_{\mathrm{out}}$), and it is  subject to the Sigmoid activation function. For the generator, $\mathrm{C}_{\mathrm{in}}$ $=$ $\mathrm{C}_{\mathrm{out}}$ $=$ $\mathrm{C}$, and $\mathrm{H} = 32$. The domain size is governed by $\widetilde{N}_h^p$, $\widetilde{N}_h^p$, which is 128 $\times$ 128 throughout this manuscript. \par

We note that we typically employ unstructured grids in the finite element solver; however, our framework in this study requires a structured data set. Thus, we pre-process our finite element data by interpolating the finite element result $p_h$ to structured grids, such as the finite element interpolation operator or cubic spline interpolation. We then replace the FOM dimension ${N}_h^p$, associated with the unstructured grid, with a pair $(\widetilde{N}_h^p, \widetilde{N}_h^p)$, associated to the structured grid. In practice, the value of $\widetilde{N}_h^p$ is often chosen independently on ${N}_h^p$. The same procedures are carried out for $\bm{u}_h$.  $\mathrm{B}$ corresponds to batch size. \par

The architecture of NLI model is presented in Table \ref{tab:unet_model1}.

\begin{table}[!ht]
\centering
\caption{Generator: NLI's detail used in this study (input and output sizes are represented by {[}$\mathrm{B}$, $\mathrm{C}$, $\widetilde{N}_h^p$, $\widetilde{N}_h^p${]}. We use hidden layers $\mathrm{H} = 32$). BN refers to batch normalization. }
\begin{tabular}{|l|c|c|c|c|}
\hline
Block                  & \multicolumn{1}{l|}{Input size} & \multicolumn{1}{l|}{Output size} & \multicolumn{1}{l|}{BN} & \multicolumn{1}{l|}{Dropout} \\ \hline
$1^{\mathrm{st}}$ convolutional layer & {[}$\mathrm{B}$, $\mathrm{C}$, 128, 128{]}            & {[}$\mathrm{B}$, 32, 128, 128{]}            &                                          &                              \\ \hline
$1^{\mathrm{st}}$ contracting block   & {[}$\mathrm{B}$, 32, 128, 128{]}           & {[}$\mathrm{B}$, 64, 64, 64{]}              & \checkmark                                        & \checkmark                            \\ \hline
$2^{\mathrm{nd}}$ contracting block   & {[}$\mathrm{B}$, 64, 64, 64{]}             & {[}$\mathrm{B}$, 128, 32, 32{]}             & \checkmark                                        & \checkmark                            \\ \hline
$3^{\mathrm{rd}}$ contracting block   & {[}$\mathrm{B}$, 128, 32, 32{]}            & {[}$\mathrm{B}$, 256, 16, 16{]}             & \checkmark                                        & \checkmark                            \\ \hline
$4^{\mathrm{th}}$ contracting block   & {[}$\mathrm{B}$, 256, 16, 16{]}            & {[}$\mathrm{B}$, 512, 8, 8{]}               & \checkmark                                        &                              \\ \hline
$5^{\mathrm{th}}$ contracting block   & {[}$\mathrm{B}$, 512, 8, 8{]}              & {[}$\mathrm{B}$, 1024, 4, 4{]}              & \checkmark                                        &                              \\ \hline
$6^{\mathrm{th}}$ contracting block   & {[}$\mathrm{B}$, 1024, 4, 4{]}             & {[}$\mathrm{B}$, 2048, 2, 2{]}              & \checkmark                                        &                              \\ \hline
$1^{\mathrm{st}}$ expanding block     & {[}$\mathrm{B}$, 2048, 2, 2{]}         & {[}$\mathrm{B}$, 1024, 4, 4{]}              & \checkmark                                        &                              \\ \hline
$2^{\mathrm{nd}}$ expanding block    & {[}$\mathrm{B}$, 1024, 4, 4{]}         & {[}$\mathrm{B}$, 512, 8, 8{]}               & \checkmark                                        &                              \\ \hline
$3^{\mathrm{rd}}$ expanding block     & {[}$\mathrm{B}$, 512, 8, 8{]}          & {[}$\mathrm{B}$, 256, 16, 16{]}             & \checkmark                                        &                              \\ \hline
$4^{\mathrm{th}}$ expanding block     & {[}$\mathrm{B}$, 256, 16, 16{]}      & {[}$\mathrm{B}$, 128, 32, 32{]}             & \checkmark                                        &                              \\ \hline
$5^{\mathrm{th}}$ expanding block     & {[}$\mathrm{B}$, 128, 32, 32{]}      & {[}$\mathrm{B}$, 64, 64, 64{]}              & \checkmark                                        &                              \\ \hline
$6^{\mathrm{th}}$ expanding block     & {[}$\mathrm{B}$, 64, 64, 64{]}       & {[}$\mathrm{B}$, 32, 128, 128{]}            & \checkmark                                        &                              \\ \hline
$2^{\mathrm{nd}}$ convolutional layer & {[}$\mathrm{B}$, 32, 128, 128{]}           & {[}$\mathrm{B}$, $\mathrm{C}$, 128, 128{]}             &                                          &                              \\ \hline
\end{tabular}
\label{tab:unet_model1}
\end{table}

\noindent
We also provide a code snippet for NLI's generator in Listing \ref{list:nli}.

\begin{lstlisting}[language=Python, caption= Illustration of the generator of NLI implementation,label={list:nli}]
class UNet(nn.Module):
    '''
    U-net Class
    A series of 6 contracting blocks followed by 6 expanding blocks to
    transform an input into the corresponding paired input, with an upfeature
    layer at the start and a downfeature layer at the end.
    Values:
        input_channels: the number of channels to expect from a given input
        output_channels: the number of channels to expect for a given output
        hidden_channels:: fixed at 32 in this manuscript
    '''
    def __init__(self, input_channels, output_channels, hidden_channels=32):
        super(UNet, self).__init__()
        self.upfeature = FeatureMapBlock(input_channels, hidden_channels)
        self.contract1 = ContractingBlock(hidden_channels, use_dropout=True)
        self.contract2 = ContractingBlock(hidden_channels * 2, use_dropout=True)
        self.contract3 = ContractingBlock(hidden_channels * 4, use_dropout=True)
        self.contract4 = ContractingBlock(hidden_channels * 8)
        self.contract5 = ContractingBlock(hidden_channels * 16)
        self.contract6 = ContractingBlock(hidden_channels * 32)

        self.expand1 = ExpandingBlock(hidden_channels * 64)
        self.expand2 = ExpandingBlock(hidden_channels * 32)
        self.expand3 = ExpandingBlock(hidden_channels * 16)
        self.expand4 = ExpandingBlock(hidden_channels * 8)
        self.expand5 = ExpandingBlock(hidden_channels * 4)
        self.expand6 = ExpandingBlock(hidden_channels * 2)
        self.downfeature = FeatureMapBlock(hidden_channels, output_channels)
        self.sigmoid = torch.nn.Sigmoid()

        self.linear1 = nn.Linear(hidden_channels * 64*2*2, hidden_channels * 64*2*2)
        self.hidden_channels = hidden_channels

def forward(self, x, y):
      '''
      Function for completing a forward pass of U-net with NLI approach:
      Given an input - permeability field, passes it through U-Net and returns the output.
      Parameters:
          x: an input of shape (batch size, channels, height, width)
          y: a time-step
      '''
      x0 = self.upfeature(x)
      x1 = self.contract1(x0)
      x2 = self.contract2(x1)
      x3 = self.contract3(x2)
      x4 = self.contract4(x3)
      x5 = self.contract5(x4)
      x6 = self.contract6(x5)

      # perform element-wise addition
      y = y.view(-1, 1)
      y = y.repeat(1, self.hidden_channels*64*2*2)
      x6 = x6.view(len(x6), -1)
      x6 = self.linear1(x6) + y
      # reshape back
      x6 = x6.view(len(x6), self.hidden_channels*64, 2, 2)

      x7 = self.expand1(x6, x5)
      x8 = self.expand2(x7, x4)
      x9 = self.expand3(x8, x3)
      x10 = self.expand4(x9, x2)
      x11 = self.expand5(x10, x1)
      x12 = self.expand6(x11, x0)
      xn = self.downfeature(x12)

      return self.sigmoid(xn)
\end{lstlisting} 

The differences between NLI and ILI are (1) NLI introduces $t^{n} \in \mathbb{T}$ to only the generator's bottleneck using element-wise addition while (2) ILI injects $t^{n} \in \mathbb{T}$ to all layers inside the generator through conditional batch normalization (CBN). The architecture of ILI model is presented in Table \ref{tab:unet_model2}.\par

\noindent
\begin{table}[!ht]
\centering
\caption{Generator: ILI's detail used in this study (input and output sizes are represented by {[}$\mathrm{B}$, $\mathrm{C}$, $\widetilde{N}_h^p$, $\widetilde{N}_h^p${]}. We use hidden layers $\mathrm{H} = 32$). CBN refers to conditional batch normalization.}
\begin{tabular}{|l|c|c|c|c|}
\hline
Block                  & \multicolumn{1}{l|}{Input size} & \multicolumn{1}{l|}{Output size} & \multicolumn{1}{l|}{CBN} & \multicolumn{1}{l|}{Dropout} \\ \hline
$1^{\mathrm{st}}$ convolutional layer & {[}$\mathrm{B}$, $\mathrm{C}$, 128, 128{]}            & {[}$\mathrm{B}$, 32, 128, 128{]}            &                                          &                              \\ \hline
$1^{\mathrm{st}}$ contracting block   & {[}$\mathrm{B}$, 32, 128, 128{]}           & {[}$\mathrm{B}$, 64, 64, 64{]}              & \checkmark                                        & \checkmark                            \\ \hline
$2^{\mathrm{nd}}$ contracting block   & {[}$\mathrm{B}$, 64, 64, 64{]}             & {[}$\mathrm{B}$, 128, 32, 32{]}             & \checkmark                                        & \checkmark                            \\ \hline
$3^{\mathrm{rd}}$ contracting block   & {[}$\mathrm{B}$, 128, 32, 32{]}            & {[}$\mathrm{B}$, 256, 16, 16{]}             & \checkmark                                        & \checkmark                            \\ \hline
$4^{\mathrm{th}}$ contracting block   & {[}$\mathrm{B}$, 256, 16, 16{]}            & {[}$\mathrm{B}$, 512, 8, 8{]}               & \checkmark                                        &                              \\ \hline
$5^{\mathrm{th}}$ contracting block   & {[}$\mathrm{B}$, 512, 8, 8{]}              & {[}$\mathrm{B}$, 1024, 4, 4{]}              & \checkmark                                        &                              \\ \hline
$6^{\mathrm{th}}$ contracting block   & {[}$\mathrm{B}$, 1024, 4, 4{]}             & {[}$\mathrm{B}$, 2048, 2, 2{]}              & \checkmark                                        &                              \\ \hline
$1^{\mathrm{st}}$ expanding block     & {[}$\mathrm{B}$, 2048, 2, 2{]}         & {[}$\mathrm{B}$, 1024, 4, 4{]}              & \checkmark                                        &                              \\ \hline
$2^{\mathrm{nd}}$ expanding block    & {[}$\mathrm{B}$, 1024, 4, 4{]}         & {[}$\mathrm{B}$, 512, 8, 8{]}               & \checkmark                                        &                              \\ \hline
$3^{\mathrm{rd}}$ expanding block     & {[}$\mathrm{B}$, 512, 8, 8{]}          & {[}$\mathrm{B}$, 256, 16, 16{]}             & \checkmark                                        &                              \\ \hline
$4^{\mathrm{th}}$ expanding block     & {[}$\mathrm{B}$, 256, 16, 16{]}      & {[}$\mathrm{B}$, 128, 32, 32{]}             & \checkmark                                        &                              \\ \hline
$5^{\mathrm{th}}$ expanding block     & {[}$\mathrm{B}$, 128, 32, 32{]}      & {[}$\mathrm{B}$, 64, 64, 64{]}              & \checkmark                                        &                              \\ \hline
$6^{\mathrm{th}}$ expanding block     & {[}$\mathrm{B}$, 64, 64, 64{]}       & {[}$\mathrm{B}$, 32, 128, 128{]}            & \checkmark                                        &                              \\ \hline
$2^{\mathrm{nd}}$ convolutional layer & {[}$\mathrm{B}$, 32, 128, 128{]}           & {[}$\mathrm{B}$, $\mathrm{C}$, 128, 128{]}             &                                          &                              \\ \hline
\end{tabular}
\label{tab:unet_model2}
\end{table}

\noindent
We also provide a code snippet for the generator of ILI in Listing \ref{list:ili}.

\begin{lstlisting}[language=Python, caption= Illustration of the generator of ILI implementation,label={list:ili}]
class UNet(nn.Module):
    '''
    U-net Class
    A series of 6 contracting blocks followed by 6 expanding blocks to
    transform an input into the corresponding paired input, with an upfeature
    layer at the start and a downfeature layer at the end.
    Values:
        input_channels: the number of channels to expect from a given input
        output_channels: the number of channels to expect for a given output
        hidden_channels:: fixed at 32 in this manuscript
    '''
    def __init__(self, input_channels, output_channels, hidden_channels=32):
        super(UNet, self).__init__()
        self.upfeature = FeatureMapBlock(input_channels, hidden_channels)
        '''
        instead of classical batch normalization, we use conditional batch
        normalization (CBN)
        '''
        self.contract1 = ContractingBlockCBN(hidden_channels, use_dropout=True)
        self.contract2 = ContractingBlockCBN(hidden_channels * 2, use_dropout=True)
        self.contract3 = ContractingBlockCBN(hidden_channels * 4, use_dropout=True)
        self.contract4 = ContractingBlockCBN(hidden_channels * 8)
        self.contract5 = ContractingBlockCBN(hidden_channels * 16)
        self.contract6 = ContractingBlockCBN(hidden_channels * 32)

        self.expand1 = ExpandingBlockCBN(hidden_channels * 64)
        self.expand2 = ExpandingBlockCBN(hidden_channels * 32)
        self.expand3 = ExpandingBlockCBN(hidden_channels * 16)
        self.expand4 = ExpandingBlockCBN(hidden_channels * 8)
        self.expand5 = ExpandingBlockCBN(hidden_channels * 4)
        self.expand6 = ExpandingBlockCBN(hidden_channels * 2)
        self.downfeature = FeatureMapBlock(hidden_channels, output_channels)
        self.sigmoid = torch.nn.Sigmoid()

        self.linear1 = nn.Linear(hidden_channels * 64*2*2, hidden_channels * 64*2*2)
        self.hidden_channels = hidden_channels

def forward(self, x, y):
      '''
      Function for completing a forward pass of U-net with NLI approach:
      Given an input - permeability field, passes it through U-Net and returns the output.
      Parameters:
          x: an input of shape (batch size, channels, height, width)
          y: a time-step
      '''
      x0 = self.upfeature(x)
      '''
      In contrast to NLI, a time-step (y) is injected into all blocks
      '''
      x1 = self.contract1(x0, y)
      x2 = self.contract2(x1, y)
      x3 = self.contract3(x2, y)
      x4 = self.contract4(x3, y)
      x5 = self.contract5(x4, y)
      x6 = self.contract6(x5, y)

      x7 = self.expand1(x6, x5, y)
      x8 = self.expand2(x7, x4, y)
      x9 = self.expand3(x8, x3, y)
      x10 = self.expand4(x9, x2, y)
      x11 = self.expand5(x10, x1, y)
      x12 = self.expand6(x11, x0, y)
      xn = self.downfeature(x12)

      return self.sigmoid(xn)
\end{lstlisting} 

\noindent
The CBN is implemented as shown in Listing \ref{list:cbn}. We also illustrate that if we have more continuous single-value parameters, we can include them in our model through CBN. 

\begin{lstlisting}[language=Python, caption= Illustration of the conditional batch normalization (CBN) implementation,label={list:cbn}]
class ConditionalBatchNorm2d(nn.Module):
    def __init__(self, num_features, dim_embed):
        super().__init__()
        self.num_features = num_features
        self.dim_embed = dim_embed
        self.bn = nn.BatchNorm2d(num_features, affine=True)
        torch.nn.init.normal_(self.bn.weight, 0.0, 0.02)
        torch.nn.init.constant_(self.bn.bias, 0)
        self.linear = ANN(1, num_features//2)
        self.embed_gamma1 = nn.Linear(num_features//2, num_features, bias=False)
        self.embed_beta1 = nn.Linear(num_features//2, num_features, bias=False)

    def forward(self, x, y):

        y = y.view(-1, 1)
        y = self.linear(y)
        out = self.bn(x)
        gamma1 = self.embed_gamma1(y).view(-1, self.num_features, 1, 1)
        beta1 = self.embed_beta1(y).view(-1, self.num_features, 1, 1)
        # possible extension if you have more than one continuous variable
        out = out + out*gamma1 + beta1 # + out*gamma2 + beta2 + out*gamma3 + beta3

        return out

class ANN(nn.Module):
    def __init__(self, input_size, dim_embed):
        super().__init__()
        hidden_size = 50

        self.fc1 = nn.Linear(input_size, hidden_size)
        self.fc2 = nn.Linear(hidden_size, hidden_size*2)
        self.fc3 = nn.Linear(hidden_size*2, hidden_size*4)
        self.fc4 = nn.Linear(hidden_size*4, hidden_size*8)
        self.fc5 = nn.Linear(hidden_size*8, dim_embed)

    def forward(self, x):

        x = x.view(x.shape[0], -1)

        x = torch.tanh(self.fc1(x))
        x = torch.tanh(self.fc2(x))
        x = torch.tanh(self.fc3(x))
        x = torch.tanh(self.fc4(x))
        x = self.fc5(x)
        return x
\end{lstlisting} 

Both NLI and ILI use the same critic presented in Table \ref{tab:disc}. The critic utilizes the contracting block, which is also used in the generator. Each contracting block of the critic uses LeakyReLU with a negative slope of 0.2 as its activation function, the $1^{\mathrm{st}}$ convolutional layer is used to map $\mathrm{C}$ $+$ conditional field to $\mathrm{H} = 8$, and it does not subject to any activation function, and and the $2^{\mathrm{nd}}$ convolutional layer is used to map $\mathrm{H} = 8$ to $\mathrm{C}$. Dissimilar to the generator where the input size is equal to output size $\widetilde{N}_h^p$, $\widetilde{N}_h^p$ $=$ 128 $\times$ 128, the discriminator input size is $\widetilde{N}_h^p$, $\widetilde{N}_h^p$ $=$ 128 $\times$ 128, while the output size is a patch matrix of size $\mathrm{PATCH_X}$ $\times$ $\mathrm{PATCH_Y}$ $=$ 8 $\times$ 8.

\noindent
\begin{table}[!ht]
\centering
\caption{Critic: NLI and ILI's detail used in this study (input size is represented by {[}$\mathrm{B}$, $\mathrm{C}$, $\widetilde{N}_h^p$, $\widetilde{N}_h^p${]}, and output size is represented by {[}$\mathrm{B}$, $\mathrm{C}$, $\mathrm{PATCH_X}$, $\mathrm{PATCH_Y}${]}. We use hidden layers $\mathrm{H} = 8$).}
\begin{tabular}{|l|c|c|c|c|c|}
\hline
Block                  & \multicolumn{1}{l|}{Input size} & \multicolumn{1}{l|}{Output size} & \multicolumn{1}{l|}{BN} \\ \hline
$1^{\mathrm{st}}$ convolutional layer & {[}$\mathrm{B}$, $\mathrm{C}$+1, 128, 128{]}            & {[}$\mathrm{B}$, 8, 128, 128{]}             &                                                                       \\ \hline
$1^{\mathrm{st}}$ contracting block   & {[}$\mathrm{B}$, 8, 128, 128{]}            & {[}$\mathrm{B}$, 16, 64, 64{]}              &                                                                      \\ \hline
$2^{\mathrm{nd}}$ contracting block   & {[}$\mathrm{B}$, 16, 64, 64{]}             & {[}$\mathrm{B}$, 32, 32, 32{]}              & \checkmark                                                                     \\ \hline
$3^{\mathrm{rd}}$ contracting block   & {[}$\mathrm{B}$, 32, 32, 32{]}             & {[}$\mathrm{B}$, 64, 16, 16{]}              & \checkmark                                                                    \\ \hline
$4^{\mathrm{th}}$ contracting block   & {[}$\mathrm{B}$, 64, 16, 16{]}             & {[}$\mathrm{B}$, 128, 8, 8{]}               & \checkmark                                                                     \\ \hline
$2^{\mathrm{nd}}$ convolutional layer & {[}$\mathrm{B}$, 128, 8, 8{]}              & {[}$\mathrm{B}$, $\mathrm{C}$, 8, 8{]}                 &                                                  \\ \hline
\end{tabular}
\label{tab:disc}
\end{table}

\noindent
The Critic implementation is as shown in Listing \ref{list:critic}.

\begin{lstlisting}[language=Python, caption= Illustration of the Critic implementation,label={list:critic}]
class Critic(nn.Module):
    '''
    Critic Class
    Structured like the contracting path of the U-net, the critic will
    output a matrix of values classifying corresponding portions of
    the output as real or fake (patch_x, patch_y - in this case is 8, 8).
    Parameters:
        input_channels: the number of input channels
        hidden_channels: is fixed at 8 throughout this manuscript
    '''
    def __init__(self, input_channels, hidden_channels=8):
        super(Discriminator, self).__init__()
        self.upfeature = FeatureMapBlock(input_channels, hidden_channels)
        self.contract1 = ContractingBlock(hidden_channels, use_bn=False)
        self.contract2 = ContractingBlock(hidden_channels * 2)
        self.contract3 = ContractingBlock(hidden_channels * 4)
        self.contract4 = ContractingBlock(hidden_channels * 8)
        self.final = nn.Conv2d(hidden_channels * 16, 1, kernel_size=1)
        self.linear1 = nn.Linear(1, 8 * 8 * 128, bias=False)
        self.linear2 = nn.Linear(1, 8 * 8)

    def forward(self, x, y, z):
        '''
        x: output from the Generator
        y: input for the Generator
        z: time-step
        '''
        x = torch.cat([x, y], axis=1)
        x0 = self.upfeature(x)
        x1 = self.contract1(x0)
        x2 = self.contract2(x1)
        x3 = self.contract3(x2)
        x4 = self.contract4(x3)

        # we introduce time-step here
        z = z.view(-1, 1)
        z = self.linear1(z)
        x4_z = x4.view(len(x4), -1)
        x4_z = torch.sum(x4_z*z, 1, keepdim=True)
        x4_z = self.linear2(x4_z)
        x4_z = x4_z.view(len(x4_z), 1, 8, 8)

        xn = self.final(x4)
        xn = xn + x4_z

        return xn
\end{lstlisting}

\end{appendices}

\bibliographystyle{plainnat}

\bibliography{references}

\end{document}